\documentclass[a4paper,12pt]{article}
\usepackage{mathrsfs,graphicx,rotating,amsmath,amsfonts,mathtools,booktabs,amssymb}
\usepackage{hyperref}\usepackage{slashed}
\usepackage[nosort]{cite}
\usepackage[table,xcdraw]{xcolor}
\hypersetup{colorlinks,bookmarksopen,bookmarksnumbered,
linkcolor=blus,pdfstartview=FitH,urlcolor=rossos,citecolor=verde}
\allowdisplaybreaks

\newcommand{\mio}[1]{}
\newcommand{\BR}{\hbox{BR}}
\newcommand{\DM}{\chi}

 \newcommand{\med}[1]{\langle #1\rangle}
 \newcommand{\doi}[1]{\href{http://dx.doi.org/#1}{[link]}}
 \newcommand{\fig}[1]{~\ref{fig:#1}}
\newcommand{\sfrac}[2]{#1/#2} 
 \newcommand{\One}{1\!\!\hbox{I}}
\newcommand{\MDM}{M_\chi} 
\newcommand{\MDMp}{M_{\chi'}} 
\newcommand{\gDM}{g_\chi}

\allowdisplaybreaks
\usepackage{multicol}
\usepackage{color}
\definecolor{rosso}{cmyk}{0,1,1,0.4}
\definecolor{rossos}{cmyk}{0,1,1,0.55}
\definecolor{rossoc}{cmyk}{0,1,1,0.2}
\definecolor{blu}{cmyk}{1,1,0,0.3}
\definecolor{blus}{cmyk}{1,1,0,0.6}
\definecolor{bluc}{cmyk}{1,1,0,0.1}
\definecolor{verde}{cmyk}{0.92,0,0.59,0.25}
\definecolor{verdec}{cmyk}{0.92,0,0.59,0.15}
\definecolor{verdes}{cmyk}{0.92,0,0.59,0.4}

\newcommand{\eq}[1]{~{\rm (\ref{eq:#1})}}

\newcommand{\MeV}{\,{\rm MeV}}
\newcommand{\GeV}{\,{\rm GeV}}
\newcommand{\TeV}{\,{\rm TeV}}

\newcommand{\Tr}{\,{\rm Tr}}

\def\circa#1{\,\raise.3ex\hbox{$#1$\kern-.75em\lower1ex\hbox{$\sim$}}\,}

\newcommand{\beq}{\begin{equation}}
\newcommand{\eeq}{\end{equation}}

\newcommand{\bea}{\begin{eqnarray}}
\newcommand{\eea}{\end{eqnarray}}
\newcommand{\be}{\begin{equation}}
\newcommand{\ee}{\end{equation}}
\font\tenrsfs=rsfs10 at 12pt
\font\sevenrsfs=rsfs7
\font\fiversfs=rsfs5
\newfam\rsfsfam
\textfont\rsfsfam=\tenrsfs
\scriptfont\rsfsfam=\sevenrsfs
\scriptscriptfont\rsfsfam=\fiversfs

\newcommand{\alf}{\alpha_{\rm eff}}

\def\circa#1{\,\raise.3ex\hbox{$#1$\kern-.75em\lower1ex\hbox{$\sim$}}\,}
\makeatletter

\def\hhref#1{\href{http://arxiv.org/abs/#1}{arXiv:#1}} 
 
\def\art{\@ifnextchar[{\eart}{\oart}}
\def\eart[#1]#2#3#4#5#6{{\rm #2}, {\em #3 \bf #4} {\rm (#6) #5} ({\em #1})}

\def\article{\@ifnextchar[{\earticle}{\oarticle}}
\def\oarticle#1#2#3#4#5#6{{\rm #1}, {\em ``#6''}, {\rm #2 #3 (#5) #4}}
\def\earticle[#1]#2#3#4#5#6#7{{\rm #2}, {\em ``#7''}, {\rm #3 #4 (#6) #5}  [\hhref{#1}]}
\def\hepart[#1]#2{{\rm #2, \em#1}}
\def\heparticle[#1]#2#3{#2, {\em ``#3''} [\hhref{#1}]}

\newcounter{alphaequation}[equation]
\def\thealphaequation{\theequation\hbox to
0.6em{\hfil\alph{alphaequation}\hfil}}
\def\eqnsystem#1{
\def\@eqnnum{{\rm (\thealphaequation)}}
\def\@@eqncr{\let\@tempa\relax \ifcase\@eqcnt \def\@tempa{& & &} \or
  \def\@tempa{& &}\or \def\@tempa{&}\fi\@tempa
  \if@eqnsw\@eqnnum\refstepcounter{alphaequation}\fi
\global\@eqnswtrue\global\@eqcnt=0\cr}
\refstepcounter{equation} \let\@currentlabel\theequation \def\@tempb{#1}
\ifx\@tempb\empty\else\label{#1}\fi
\refstepcounter{alphaequation}
\let\@currentlabel\thealphaequation
\global\@eqnswtrue\global\@eqcnt=0 \tabskip\@centering\let\\=\@eqncr
$$\halign to \displaywidth\bgroup \@eqnsel\hskip\@centering
$\displaystyle\tabskip\z@{##}$&\global\@eqcnt\@ne
\hskip2\arraycolsep\hfil${##}$\hfil& \global\@eqcnt\tw@\hskip2\arraycolsep
$\displaystyle\tabskip\z@{##}$\hfil
\tabskip\@centering&\llap{##}\tabskip\z@\cr}
\def\endeqnsystem{\@@eqncr\egroup$$\global\@ignoretrue} \makeatother

\newcommand{\eV}{\,{\rm eV}}

\newcommand{\SU}{\,{\rm SU}}

\begin{document}
\centerline{CERN-TH-2017-030  \hfill IFUP-TH/2017}

\vspace{1cm}

\begin{center}
{\LARGE\bf\color{rossos} Cosmological Implications\\[3mm] of Dark Matter Bound States}\\

\bigskip
\bigskip
\bigskip

{\bf Andrea Mitridate$^{a}$, Michele Redi$^{b}$, Juri Smirnov$^{b}$ and Alessandro Strumia$^{c,d}$ }  
\\[7mm]

{\it $^a$ Scuola Normale Superiore, Piazza dei Cavalieri 7, 56126, Pisa, Italy}\\[1mm]
{\it $^b$ INFN, Sezione di Firenze, and Department of Physics and Astronomy, University of Florence,
Via G. Sansone 1, 50019 Sesto Fiorentino, Italy}\\[1mm]
{\it $^c$ Dipartimento di Fisica dell'Universit{\`a} di Pisa and INFN, Italy}\\[1mm]
{\it $^d$ CERN, Theory Division, Geneva, Switzerland}\\[1mm]

\begin{figure}[t]
\begin{center}
\label{fig:gluinos}
\end{center}
\end{figure}

\vspace{2cm}
{\large\bf\color{blus} Abstract}
\begin{quote}\large
We present generic formul\ae{} for computing how 
Sommerfeld corrections together with
bound-state formation affect the thermal abundance of Dark Matter
with non-abelian gauge interactions.
We consider DM as a fermion 3plet (wino) or 5plet under $\SU(2)_L$.
In the latter case bound states raise  the DM mass required to reproduce the cosmological DM abundance to 14 TeV
and give new indirect detection signals such as (for this mass) a dominant $\gamma$-line around 85 GeV.
Furthermore, we consider DM co-annihilating with a colored particle, such as a squark or a gluino,
finding that bound state effects are especially relevant in the latter case.
\end{quote}

\bigskip
\bigskip
\bigskip

\thispagestyle{empty}
\end{center}
\begin{quote}
{\large\noindent\color{blus} 
}

\end{quote}

\newpage

\tableofcontents

\newpage

\section{Introduction}
The hypothesis that Dark Matter (DM) is a thermal relic of a weakly interacting particle 
allows to use the cosmological DM abundance
$\Omega_{\rm DM} h^2 = 0.119\pm 0.002$~\cite{1502.01589} to derive information on the DM mass. The latter
gets fixed in theories with no extra free parameters such as Minimal Dark Matter~\cite{hep-ph/0512090} and, 
even allowing for extra production mechanisms, one obtains interesting constraints.
Thus it is crucial to compute thermal freeze-out abundance accurately. 
For this purpose we will study non-relativistic scatterings among particles with mass $M$ charged under a gauge group $G$ with gauge coupling $g$ and mediated by vectors $V$ with mass $M_V$.
Those get significantly suppressed or enhanced by Coulomb-like forces if $M_V < \alpha M$, where $\alpha=g^2/4\pi$. The relevance of this Sommerfeld effect for annihilations of
Dark Matter particles has been recognised long time ago~\cite{HisanoCosmo,Cirelli:2007xd,Khlopov}:
$\sigma v$ roughly grows as $ v_{\rm max}/ v$ in the range  $v_{\rm min}< v <v_{\rm max}$ where
$ v_{\rm max} \approx g^2/4$ and $v_{\rm min}\approx M_V/M$
(or smaller if one bound state happens to have zero energy).
Thereby, the Sommerfeld effect is relevant at  temperatures $T\circa{<}\alpha^2M$.

Recent literature~\cite{1407.4121,Petraki,Ellis:2015vaa,Slatyer,1611.08133} (see \cite{0812.0559} for earlier work) 
recognised that a second related phenomenon is important too: formation of bound states $B$ of two Dark Matter particles with binding energy of order $\alpha^2 M$,
through processes analogous to the  formation of hydrogen at recombination.
The two DM particles within the bound state annihilate with rate $\Gamma_{\rm ann}\sim \alpha^5 M$.

This effect has been considered only more recently, possibly for the following reason.
Naively one  expects that at $T \circa{>} \alpha^2 M$ 
scatterings with the thermal plasma rapidly break the bound states before they can annihilate,
such that bound state formation would be irrelevant at the temperature $T \approx M/25$ of Dark Matter decoupling
(unless $\alpha$ is very large).
The above argument misses a feature of non-relativistic interactions: the rate $\Gamma_{\rm break}$ for
breaking the bound state is suppressed by $\alpha^5$ at $T \circa{<}   \alpha M$, when
particles $V$ in the thermal plasma have a wave-length smaller than the size $a_0 \sim 1/\alpha M$ of the bound state.
The thermal rate for breaking the bound state, $\Gamma_{\rm break}$, can then be comparable to $\Gamma_{\rm ann}$.
Moreover, at low enough velocities, the cross-section for bound states formation is parametrically
comparable to the
Sommerfeld-corrected annihilation cross-section.

So far, bound-state effects have mostly been considered in models of Dark Matter charged under a speculative abelian extra `dark force', see~\cite{DF} and references therein.
We study how bound state formation affects annihilations of DM particles with SM gauge interactions, $\alpha \sim \alpha_{1,2}$,
as well as co-annihilations with colored particles, $\alpha \sim \alpha_3$.
We will find that bound state formation indeed gives significant effects.

\medskip

The paper is structured as follows.
In section~\ref{setup} we show how the system of Boltzmann equations for DM freeze-out
can be reduced to a single equation with an effective annihilation cross section that takes into account
Sommerfeld corrections and bound state formation.
In section~\ref{Sommerfeldsec} we review how the Sommerfeld correction can be computed for
non-abelian gauge interactions, and how the effect of non-zero vector masses can be approximated analytically.
In section~\ref{bound} we summarise the basic formul\ae{} for bound state formation,
showing how the effects of non-abelian gauge interactions can be encoded into Clebsh-Gordon-like factors,
and how the main effect of massive vectors is kinematical.
In section~\ref{sec:decays} we provide formul\ae{} which describe the main properties of the bound states, such as
annihilation rates and decay rates.
All these quantities are needed at finite temperature: 
in section~\ref{Thermal} we discuss the issue of thermal corrections,
showing that breaking of gauge interactions lead to the loss of quantum coherence.
Finally, in section~\ref{Applications} we perform concrete computations in interesting models
of Dark Matter charged under $\SU(2)_L$ (a wino triplet, a quintuplet) and of co-annihilation with particles charged under $\SU(3)_c$ (squarks and gluinos).
We find that bound state effects can be sizeable, as summarized in the conclusion, section~\ref{concl}.


\section{Setup of the computation}\label{setup}
We assume that DM $\chi_i$ lies in the representation $R$ (if real) or $R\oplus \overline R$ (if $R$ is complex)
of a gauge group $G$ with gauge coupling $g$.
We define $\alpha = g^2/4\pi$ and $g_\chi$ as the number of degrees of freedom of the DM system.
In practice we will consider the following cases:
\begin{enumerate}
\item DM is the neutral component of a triplet under electroweak $\SU(2)_L$ with zero hypercharge e.g.\ a supersymmetric wino.
Then $g_\chi=6$.

\item DM is the neutral component of a quintuplet under electroweak $\SU(2)_L$ with zero hypercharge.
Then $g_\chi=10$.
\item DM is a singlet that co-annihilates with squarks, that form a $3$ under color $\SU(3)_c$.

\item DM is a singlet that co-annihilates with gluinos, that form a $8$ under color $\SU(3)_c$.
\end{enumerate}
We want to compute the DM freeze-out that happens around $T\sim \MDM/25$ and below,
when various non-relativistic effects give non-perturbative corrections:
the Sommerfeld enhancement and formation of bound states of two DM particles.
This is done by solving cosmological Boltzmann equations, that contain the various particle-physics that we will compute in the next sections.

\subsection{Boltzmann equations}
We show how the system of Boltzmann equations for the DM number density $n_{\rm DM}$
and for the number density $n_I$ of
the various bound states can be reduced to a single equation for the DM density
with an effective DM annihilation cross section.
We define an index $I$ that identifies each DM bound state, and that
collectively denotes its various quantum numbers:
angular momentum, spin, gauge group representation, etc.

\smallskip

The Boltzmann equation for the total DM density is
\beq \label{eq:YDM}
sHz \frac{dY_{\rm DM}}{dz} = -2\gamma_{\rm ann}
\bigg[\frac{Y_{\rm DM}^2}{Y_{\rm DM}^{\rm eq2} }-1\bigg]-2\sum_{I}
\gamma_I  \bigg[\frac{Y_{\rm DM}^2}{Y_{\rm DM}^{\rm eq2}} -\frac{Y_I}{Y_I^{\rm eq}}\bigg]
\eeq
where $Y_{\rm DM} = n_{\rm DM}/s$, $s$ is the entropy density, $z = \MDM/T$.
We define as $n^{\rm eq}$ and $Y^{\rm eq}$ the value that each $n$ or $Y$ would have in thermal equilibrium, and
$\gamma$ is the space-time density of interactions in thermal equilibrium,
connected to cross sections as summarized in~\cite{Cirelli:2009uv}.
The first term describes DM DM annihilations to SM particles; the extra term describes formation of bound state $I$.

We next need  the Boltzmann equation for the number density of bound state $I$, $n_I(t)$:
\beq  \label{eq:nI}
\frac{ \dot n_I+ 3H n_I}{n_I^{\rm eq}}=
 \med{ \Gamma_{I\rm break} } 
\bigg[\frac{n_{\rm DM}^2}{n_{\rm DM}^{\rm eq2}} - \frac{n_I}{n_I^{\rm eq}}\bigg]
+\med{\Gamma_{I\rm ann}} 
\bigg[1-\frac{n_I}{n_I^{\rm eq}}  \bigg]+\sum_J
\med{\Gamma_{I\to J}} 
\bigg[\frac{n_J}{n_J^{\rm eq}}-\frac{n_I}{n_I^{\rm eq}}  \bigg]  .
\eeq
The first term accounts for formation from DM DM annihilations and breaking:
$ \med{ \Gamma_{I\rm break} } $ is the thermal average of the breaking rate of bound state $I$ due to its collisions with the plasma.
The second term contains $ \med{ \Gamma_{I\rm ann} } $, which is the thermal average of the decay rate of the bound 
state $I$ into SM particles, due to annihilation of its DM components.
The third term describes decays to lower bound states $J$ or from higher states $J$, as well as the inverse excitation processes.
They are both accounted in a single term if we define $\Gamma_{J\to I} = - \Gamma_{I\to J}$.
For decays, the thermal average of the Lorentz dilatation factor of a particle with total mass $M$ gives
$\med{\Gamma} = \Gamma K_1(M/T)/K_2(M/T)$, which equals to the decay width at rest $\Gamma$
in the non-relativistic limit $T\ll M$.
The thermal rate for breaking has a different dependence on $T$.
In the models we consider at least some decay or annihilation rates is much faster than the Hubble rate, $\Gamma \gg H$.
Therefore, the left-handed side of eq.\eq{nI} can be neglected, 
and the system of differential equations reduces to a system of linear equations that determine the various $n_I/n_I^{\rm eq}$. 
This can be shown formally by rewriting eq.\eq{nI} for $n_I(t)$ into an
equivalent Boltzmann equation for $Y_I(z)$
\beq  \label{eq:nI2}
sHz\frac{dY_I}{dz} =n_I^{\rm eq}\bigg\{
 \med{ \Gamma_{I\rm break} } 
\bigg[\frac{Y_{\rm DM}^2}{Y_{\rm DM}^{\rm eq2}} - \frac{Y_I}{Y_I^{\rm eq}}\bigg]
+\med{\Gamma_{I\rm ann}} 
\bigg[1-\frac{Y_I}{Y_I^{\rm eq}}  \bigg]+\sum_J
\med{\Gamma_{I\to J}} 
\bigg[\frac{Y_J}{Y_J^{\rm eq}}-\frac{Y_I}{Y_I^{\rm eq}}  \bigg]  \bigg\}.
\eeq
Inserting the values of $n_I$ or $Y_I$ into eq.\eq{YDM}, it becomes one differential equation for the DM abundance with
an effective cross section
\beq \label{eq:YDMeff}
sHz \frac{dY_{\rm DM}}{dz} = -2\gamma_{\rm eff}
\bigg[\frac{Y_{\rm DM}^2}{Y_{\rm DM}^{\rm eq2} }-1\bigg].
\eeq
For example, in the case of a single bound state $I=1$ one finds
\beq\label{eq:B1}
\gamma_{\rm eff} = \gamma_{\rm ann} + \gamma_1 {\rm BR}_1,\qquad
{\rm BR}_1=
 \frac{\med{\Gamma_{1\rm ann}}}{ \med{\Gamma_{1\rm ann}+\Gamma_{1\rm break}}} .\eeq
Namely, the rate of DM DM annihilations into the bound state gets multiplied by its branching ratio into SM particles.\footnote{
In the case of two bound states 1 and 2 one finds
\beq\label{eq:2bs}
\gamma_{\rm eff} = \gamma_{{\rm DM}\to{\rm SM} } + \frac{\gamma_1 
(\med{\Gamma_{\rm 1ann}}\med{\Gamma_{2}} + \med{\Gamma_{12}}\med{\Gamma_{1\rm ann}+\Gamma_{2\rm ann}})
+\gamma_2
(\med{\Gamma_{\rm 2ann}}\med{\Gamma_{1}} + \med{\Gamma_{12}}\med{\Gamma_{1\rm ann}+\Gamma_{2\rm ann}})
}
{ \med{\Gamma_1} \med{\Gamma_2 }+ \med{\Gamma_{12}}\med{\Gamma_1+\Gamma_2}}
\eeq
where $\Gamma_I \equiv \Gamma_{I\rm ann} + \Gamma_{I\rm break}$.}
The breaking rate $\Gamma_{I\rm break}$ is related to the 
space-time density formation rate $\gamma_I$ by the
Milne relation 
\beq
\gamma_I = n_I^{\rm eq}  \med{ \Gamma_{I\rm break} } .\eeq 
It is derived
taking into account that 2 DM particles disappear whenever a DM-DM bound state forms, such that 
$Y_{\rm DM}+Y_I/2$ is conserved by this process,
and by comparing eq.\eq{nI2} with eq.\eq{YDM}.

Next, the space-time densities $\gamma$ for DM-DM process can be
written in the usual way in terms of the cross sections $\sigma v_{\rm rel}$, averaged over all DM components.\footnote{If DM is a real particle (e.g.\ a Majorana fermion) this is the usual definition of a cross section.
If DM is a complex particle (e.g.\ a Dirac fermion) with no asymmetry,
the average over the 4 possible  initial states is
$\sigma \equiv \frac{1}{4}(2\sigma_{\chi\overline\chi} + 
\sigma_{\chi\chi} + \sigma_{\overline\chi\overline\chi})$.
In many models only $\chi\overline\chi$ annihilations are present, so that
$\sigma =  \frac12 \sigma_{\chi\overline\chi}$.}
In the  non-relativistic limit one has 
\beq 2\gamma \stackrel{T\ll \MDM}{\simeq} (n_{\rm DM}^{\rm eq})^2 \med{\sigma v_{\rm rel}}.\eeq
The cosmological DM abundance is approximatively
reproduced if $\med{\sigma_{\rm eff} v_{\rm rel}}$ equals to the value in eq.\eq{sigmavthermal}.
More precisely,  the Boltzmann equation for the DM abundance can be written in the final form
\beq  \frac{dY_{\rm DM}}{dz} = - \frac{\med{\sigma_{\rm eff} v_{\rm rel}} s} {Hz} (Y_{\rm DM}^2 - Y_{\rm DM}^{\rm eq2})=
- \frac{\lambda \, S(z)} {z^2} (Y_{\rm DM}^2 - Y_{\rm DM}^{\rm eq2}),
\eeq
where $S$ is the tempeature-dependent correction due to higher order effects
(Sommerfeld enhancement, bound-state formation, \ldots)
with respect to a reference cross section $\sigma_0$ computed at tree level in $s$-wave 
\beq S(z) = \frac{\med{\sigma_{\rm eff} v_{\rm rel}}}{\sigma_0} ,
\qquad
 \lambda = \left.\frac{\sigma_0 s}{H}\right|_{T=\MDM} = 
 \sqrt{\frac{g_{\rm SM}\pi}{45}} \sigma_0 M_{\rm Pl} \MDM
\eeq
where $g_{\rm SM}$ is the number of degrees of freedom in thermal equilibrium at $T=\MDM$
($g_{\rm SM}=106.75$ at $T\gg M_Z$) and $M_{\rm Pl}=G_N^{-1/2}=1.22 \times 10^{19}$ GeV.
In the non-relativistic limit the Milne relation becomes 
\beq\med{ \Gamma_{I\rm break}}= \frac{g_\chi^2}{g_I} \frac{(\MDM T)^{3/2}}{16\pi^{3/2}} e^{-E_{B_I}/T}
\med{\sigma_{I} v_{\rm rel}}
\label{eq:gammabreak}
\eeq
where $E_{B_I}>0$ is the binding energy of the bound state under consideration, $g_I$ is the number of its degrees of freedom,
and $\langle \sigma_{I} v_{\rm rel} \rangle$ is the thermal average of the cross section for bound-state formation
(computed in section~\ref{bound}).
The branching ratio in eq.\eq{B1} approaches 1 at small enough temperature.
For a single bound state one has the explicit result
\begin{align}
S (z) = S_{ \rm ann}(z)  + \left[\frac{\sigma_0}{\langle \sigma_{I} v_{\rm rel} \rangle} + \frac{g_{\chi}^2 \sigma_0\,\MDM^3 }{2 g_I\, \Gamma_{\rm ann}} \left(\frac{1}{4 \pi z}\right)^{3/2} e^{- z\, E_{B_I}/\MDM}  \right]^{-1}
\end{align}
where  $S_{\rm ann}$ is the Sommerfeld correction to the annihilation cross section (computed in section~\ref{Sommerfeldsec}),
and the second term is the contribution from the bound state $I$.
Its effect is sizeable if $\sigma_{I} $, $E_{B_I}$ and $\Gamma_{I\rm ann}$ are large.

\bigskip

The single Boltzmann equation can be integrated to obtain the final dark matter abundance $Y_{\rm DM}({\infty}) $. 
Extending the boundary layer method~\cite{Bender} to a generic $S(z)$
gives the approximated solution
\begin{align}
Y_{\rm DM}({\infty}) =  \frac{1}{ \lambda } \left(  \int_{z_f}^\infty \frac{ S(z)}{z^2} dz   +   \frac{   S (z_f)}{z_f^2} \right)^{-1} \,,
\end{align}
with the freeze out epoch $z= z_f$ given by
\begin{align}
z_f&  = \ln{\left( \frac{2  g_\chi  S(z_f) \lambda}{  (2  \pi z_f)^{3/2}} \right)}\,.
\end{align}
This approximations is accurate when, as in the situation under study, there are extra annihilations
at later times, as encoded in the factor $S$.
The relic DM density is
\beq \Omega_{\rm DM} \equiv \frac{\rho_{\rm DM}}{\rho_{\rm cr}} =\frac{s_0Y_{\rm DM}(\infty)  \MDM}{3H_0^2/8\pi G_N} =
\frac{0.110}{h^2} \times  \frac{Y_{\rm DM}(\infty)  \MDM}{0.40\eV}.
\eeq
As well known, assuming that the effective (co)annihilation cross section averaged over all DM components
is approximatively constant,
thermal freeze-out reproduces the
observed cosmological DM abundance  when it  equals
\be\label{eq:sigmavthermal}
\langle\sigma_{\rm eff} v_{\rm rel}\rangle_{\rm cosmo}\approx  2.2 \times 10^{-26}\, \frac{\textrm{cm}^3}{\textrm{s}} =
\frac{1}{(23\TeV)^2}
\ee
at $T\approx \MDM/25$.
Here  $v_{\rm rel} \ll 1$ is the DM velocity in the center-of-mass frame.
In the next sections we describe how $\sigma v_{\rm rel}$ can be computed.

\section{Sommerfeld enhancement}\label{Sommerfeldsec}

\subsection{DM annihilation at tree level}\label{tree}
The tree-level (co)annihilation cross section of DM particles into SM particles
can be readily computed.
We consider two main class of models.
In both cases we assume that the DM mass is much heavier than all SM particles.
A posteriori, this will be consistent with the DM cosmological abundance.

\medskip

First, we assume that DM is the neutral component of a fermionic $n$-plet of $\SU(2)_L$
with hypercharge $Y=0$ and mass $\MDM$. 
The $s$-wave annihilation cross section into SM vectors, fermions and Higgses is~\cite{Cirelli:2007xd} 
\begin{equation}\label{eq:MDMtree}
\sigma v_{\rm rel} =
 \frac {g_2^4(2 n^4 +17 n^2-19)}{256\pi \gDM\MDM^2}=
\frac{\pi\alpha_2^2}{\MDM^2}
 \left\{\begin{array}{ll}
 37/12 & n=3\\
 207/20 & n=5
 \end{array}\right. 
\end{equation}
where $\gDM =2n$ is the number of degrees of freedom of the DM multiplet.
The $p$-wave contribution is suppressed by an extra $v_{\rm rel}^2$ factor.
Similar formul\ae{} apply for fermions with $Y\neq 0$ and for a degenerate scalar multiplet~\cite{Cirelli:2007xd}.
Related interesting models have been proposed along similar lines~\cite{Nardecchia}.

\medskip

Next, we consider co-annihilations of a DM particle $\chi$ with $g_\chi$ degrees of freedom
with a colored state
 $\chi'$  in the representation $R$ of $\SU(3)_c$
 and mass $M_{\chi'} = \MDM + \Delta M$.
 In supersymmetric models $\chi$ can be a neutralino and $\chi'$ can be the gluino or a squark. 
  Assuming that  co-annihilations  are dominant one has an effective cross-section~\cite{1402.6287}
 \be\label{eq:sigmav}
 \sigma v_{\rm rel}=
\sigma(\chi'\chi'\to \textrm{SM particles})v_{\rm rel}
\times
\left[ 1+ \frac{g_\chi}{g_{\chi'} }\frac{\exp(\Delta M/T)}{(1+\Delta M/\MDM)^{3/2}}
\right]^{-2}\ .\eeq
Assuming that $\chi'$ lies in the representation $R$ of color $\SU(3)_c$ one has the $s$-wave cross sections~\cite{1402.6287}
\begin{eqnsystem}{sys:col}  \label{eq:ann2gluons}
\sigma(\chi'\chi'\to g g )v_{\rm rel} &=&\frac{2d_R C^2_R -12 T_R}{ g_{\chi^\prime}d_R}\frac{\pi \alpha_3^2}{M_{\chi^\prime}^2} \,, \\
\sigma(\chi'\chi'\to q\overline q)v_{\rm rel}&=&\frac{48 T_R}{ g_{\chi^\prime}d_R}\frac{\pi \alpha_3^2}{M_{\chi^\prime}^2}\times\left\{
\begin{array}{ll}
1 & \hbox{if $\chi'$ is a fermion}\\
0 & \hbox{if $\chi'$ is a boson}
\end{array}\right.
\,.
\end{eqnsystem}
where we summed over all SM quarks and
$d_3=3$, $T_3 = 1/2$, $C_3=4/3$;
$d_8=8$, $T_8=C_8=3$, 
$C_{10}=C_{\overline{10}}=6$, $C_{27}=8$, etc.
The number of  degrees of freedom of $\chi'$ is  $g_{\chi'}=6$ for a scalar triplet,
8 for a scalar octet, 12 for a fermion triplet,  16 for a fermion octet.

\medskip

As discussed in the next sections, all these tree-level cross sections get significantly affected by 
Sommerfeld corrections and by bound-state formation
due to SM gauge interactions.

\subsection{Sommerfeld corrections}\label{Sommerfeld}
We  consider an arbitrary gauge group with a common vector mass $M_V$.
Non-abelian interactions among particles in the representations $R$ and $R'$ give rise to the non-relativistic potential
\begin{equation}
V= \alpha  \frac{ e^{-M_V r}}{r} \sum_a T_R^a\otimes  T_{R'}^a
\end{equation}
which is a matrix, if written in $R,R'$ components.
As long as the group is unbroken,
its algebra allows to decompose the processes into effectively abelian sub-sectors,
$R\otimes R'= \sum_J J$, as
\begin{equation}
V=\alpha \frac {e^{-M_V r}}{2 r} \left[\sum_J C_J  \One_J- C_R  \One_R - C_{R} \One_{R'}\right].
\end{equation}
In each sub-sector one gets an effective abelian-like potential described by a numerical constant $\lambda_J$.
\begin{equation}\label{eq:VQ}
V_J =-\alf\frac {e^{-M_V r}}{r},\qquad \alf = \lambda_J \alpha, \qquad
\lambda_J = \frac{C_R+C_{R'}-C_J}{2} \eeq
such that $\alf>0$ and $\lambda_J>0$ for an attractive channel $J$. 

\medskip

We specialise to the two classes of models considered in section~\ref{tree}.

Isospin $\SU(2)_L$ is broken,  and gets restored by thermal effects at $T \circa{>}155\GeV$,
where degenerate vector thermal masses $M_V$ respect the group decomposition.
The Casimir of the $\SU(2)_L$ irreducible representations with dimension $n$ is $C_n=(n^2-1)/4$.
A two-body state decomposes as $n\otimes\overline n=1\oplus 3\oplus\ldots\oplus
2n-1$. The potential is
$V = (I^2+1-2n^2)\alpha_2/8r$
within the two-body sector with isospin $I$.
The most attractive channel is the singlet $I=1$: $V=-2\alpha_2/r$ for $n=3$
($\alf=0.066$),
$V=-6\alpha_2/r$ for $n=5$
($\alf=0.2$).

Color is unbroken. The Casimirs $C_R$ of SU(3) irreducible representations have been listed above,
such that the singlet state has $V=-4\alpha_3/3r$ if made of $3\otimes \overline 3$
($\alf=0.13$)
and $V=-3\alpha_3/r$ if made of $8\otimes 8$
($\alf=0.3$).

\bigskip

The Sommerfeld correction can be computed from the distortion of the wave function of the initial state.
In the center of mass frame of the incoming two 2 fermions, the stationary Schroedinger equation is
\begin{equation}
\label{eq:Schro}
-\frac{\nabla^2\psi}{\MDM} +  V \psi = E \psi.
\end{equation}
As usual we can decompose the wave function in states of given orbital angular momentum
\beq
\psi(r,\theta,\varphi) = R_\ell (r)Y_\ell^m(\theta,\varphi)=
\frac{u_\ell(r)}{r} Y_\ell^m(\theta,\varphi)\eeq
where $Y_\ell^m$ are spherical harmonics and the radial wave function $u_\ell(r)$  satisfies
\beq
-\frac{u'' _\ell}{\MDM}+\left[V +\frac{\ell(\ell+1)}{\MDM r^2}\right]u_\ell =Eu_\ell.
\eeq
The Schroedinger equation admits discrete solutions with negative energy and
continuum solutions with 
$E = \MDM v_{\rm rel}^2/4$
equal to the kinetic energy of the two DM particles in the center-of-mass
frame, where each DM particle  has velocity $\beta$,
such that their relative velocity is $v_{\rm rel}=2\beta$.
For identical particles, one must only consider a wave function (anti)symmetric under their exchange.

The deflection of the initial wave function from a plane wave leads to the Sommerfeld enhancement. 
For $s$-wave annihilation,\footnote{The Sommerfeld enhancement also affects $p$-wave cross sections, which remain subleading~\cite{Iengo,Cassel:2009wt}.} 
the Sommerfeld factor that enhances the tree-level cross section can be computed as
$S = |u(\infty)/u(0)|^2$
where $u$ has outgoing boundary condition $u'(\infty)/u(\infty) \simeq i\MDM v_{\rm rel}/2$.
For the potential of eq.\eq{VQ} and  $s$-wave scattering one gets
\beq  \label{eq:S0}
S = \frac{2\pi \alf/v_{\rm rel}}{1-e^{-2\pi \alf/v_{\rm rel}}} \qquad
\hbox{for $M_V=0$.}\eeq
In the case of a massive vector, an analytic solution is obtained approximating the Yukawa potential 
with a Hulthen potential
\beq\label{eq:Hulthen}
 \frac{e^{- M_V r}}{r}\approx  \frac{  \kappa\, M_V e^{- \kappa M_V r }}{1- e^{-\kappa\, M_V r}}\,.
\eeq
This potential approximates the Yukawa behaviour best if $\kappa$ is chosen as 
 $\kappa \approx 1.74$. 
The  Sommerfeld factor that enhances an $s$-wave cross section is~\cite{Cassel:2009wt}
\begin{align}\label{eq:Som}
S = \frac{2 \pi  \alf  \sinh \left(\sfrac{\pi  \MDM  v_{\text{rel}}}{\kappa 
   M_V}\right)}{v_{\text{rel}} \left(\cosh \left(\sfrac{\pi  \MDM
   v_{\text{rel}}}{\kappa  M_V}\right)-\cosh \left(\sfrac{\pi  \MDM
   v_{\text{rel}} \sqrt{1-\sfrac{4 \alf  \kappa  M_V}{\MDM
   v_{\text{rel}}^2}}}{\kappa  M_V}\right)\right)}\,.
\end{align}
This expression reduces to the Coulomb result of eq.~(\ref{eq:S0}) in the limit of vanishing vector mass $M_V$. 
$S$ is resonantly enhanced when $\MDM = \kappa n^2 M_V/\alf$ for integer $n$,
which corresponds to a zero-energy bound state, as discussed in section~\ref{bound}.
$S$ depends only on $\alf/v_{\rm rel}$ and on $y\equiv \kappa M_V/\MDM \alf$;
its thermal average $\med {S}$
depends only on $\alf \sqrt{z}$ and $y$, where $z=\MDM/T$.
At small velocities, $ v_{\rm rel}\ll M_V/\MDM$ as relevant for indirect detection, the formula above reduces to
\begin{equation}
S \stackrel{{v_{\rm rel}\rightarrow 0}}\simeq\frac {2 \pi^2 \alf \MDM}{\kappa M_V } 
\left(1- \cos 2\pi \sqrt{\frac {\alf \MDM}{\kappa M_V}}\right)^{-1}
\end{equation}
producing a significant enhancement if $ \alf \MDM/M_V\circa{>}1$.

\begin{figure}[t]
\begin{center}
\includegraphics[width=0.5\textwidth]{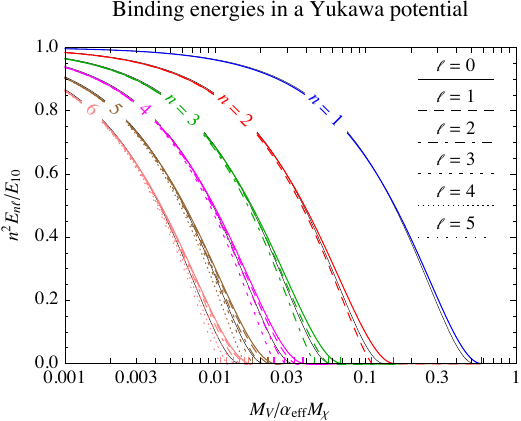}
\caption{\label{fig:YukawaEbinding}\em  Energies of bound states in a Yukawa potential (colored curves)
compared to the Hulthen approximation
with $\kappa=1.9$ (black continuous curves).}
\end{center}
\end{figure}

\section{Bound state formation}\label{bound}

\subsection{Binding energies}
As well known,
an infinity of bound states with quantum number $n=1,2,\ldots$ exist in a Coulomb potential 
$V = - \alf/r$ with any $\alf$:
the binding energies $E_B$ are $E_{n\ell} = \alf^2 \MDM/4n^2 $ and do not depend on the angular momentum $\ell$;
their  wave functions normalized to unity 
$\psi_{n\ell m} (r,\theta,\varphi)  =R_{n\ell}(r) Y_{\ell m}(\theta,\varphi)$ 
are summarized in eq.\eq{psiCoulomb} in the appendix.
In particular,
$ \psi_{100}(r,\theta,\varphi) = \sfrac{e^{-r/a_0}}{\sqrt{\pi a_0^3}}$
for the ground state, where $a_0= 2/\alf \MDM$ is the Bohr radius.


\smallskip

A Yukawa potential $-\alf e^{-M_V r}/r$ allows a finite number of
bound states if the Yukawa screening length, $1/M_V$, is larger than the Bohr radius:
$M_V \circa{<} \alf \MDM$.
Formation of a bound state via emission of a vector is kinematically possible if 
the binding energy $\sim \alf^2 \MDM$ plus the kinetic energy  $ \MDM v_{\rm rel}^2/4$ is larger than the mass of the emitted vector:
$M_V \circa{<} (\alf^2 + v_{\rm rel}^2)\MDM$~\cite{Petraki}.

\smallskip

The binding energies in a Yukawa potential can be exactly computed at first order in $M_V$ by expanding
$V=-\alf \exp(-M_V r)/r \simeq-\alf (1/r- M_V)$, finding
\begin{equation}\label{eq:EBsmallMV}
 E_{n\ell}\simeq  \frac{\alpha_\text{eff}^2 M_\chi}{4 n^2}   -\alf M_V + {\cal O}(M_V^2).
\end{equation}
The relative  correction becomes of order unity for $M_V \sim \alf \MDM$ where the Coulomb approximation is unreliable.
The shift in energy is equal for ground state and excited levels so that the Coulomb approximation fails earlier for the latter ones.

\smallskip

Fig.\fig{YukawaEbinding} shows numerical results for the binding energies, obtained by computing the matrix elements
of the Yukawa potential in the basis of eq.\eq{psiCoulomb} and diagonalising the resulting matrix
in each sector with given $\ell$, see also~\cite{Yukawa,Slatyer}.
Analytic expressions for the binding energies are obtained by approximating
the Yukawa potential with the  Hulthen potential of eq.\eq{Hulthen}, where $\kappa$ is an arbitrary order one constant.
For states with  $\ell=0$ one has
\beq\label{eq:EBHul}
E_{n0} = \frac{\alpha_\text{eff}^2 M_\chi}{4 n^2}  \left[  1-  n^2 y\right]^2\qquad\hbox{where}\qquad
 y \equiv \frac{\kappa M_V}{\alf\MDM}
\eeq
which  reproduces eq.\eq{EBsmallMV} at leading order in $M_V$ for $\kappa=2$.
The bound state exists only when the term in the squared parenthesis is positive,
namely for $\MDM \ge \kappa n^2 M_V/\alf$.
Fig.\fig{YukawaEbinding} shows that setting  $\kappa\approx 1.90$ better reproduces the generic situation,
while $\kappa \approx 1.74$ better reproduces the 
critical value at which the  special $n=1$ bound state first forms. 
Bound states with angular momentum
$\ell>0$ have different energy from the corresponding state with $\ell=0$
only if the Yukawa potential deviates significantly from its Coulomb limit,
namely if the second term in the parenthesis is of order one.
Analytic solutions are only available making extra simplifications.
A comparison with numerical results suggests a relatively minor correction of the form
\beq\label{eq:EboundYuk}
E_{n\ell} \approx  \frac{\alpha_\text{eff}^2 M_\chi}{4 n^2}  
 \bigg[  1-   n^2 y-0.53 n^2 y^2 \ell (\ell+1) \bigg]^2,\qquad {\kappa=1.74} .
\eeq
The wave functions for free and bound states, in a Coulomb or Hulthen potential, will be needed later and
are listed in the appendix.

\begin{figure}[t]
\begin{center}
\includegraphics[width=0.85\textwidth]{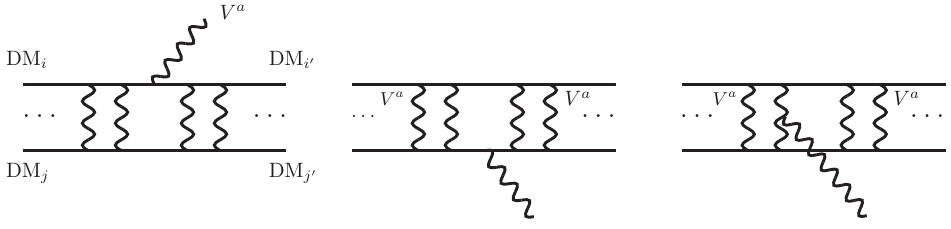}
\caption{\label{fig:bsf}\em  
Diagrams relevant for bound state formation. 
The first two diagrams give the first two terms of eq.~\eqref{eq:HI}. The third diagrams, which is peculiar of non-abelian interactions, gives rise to the last term.}
\end{center}
\end{figure}

\subsection{Bound state formation}
\label{sec:bsf}

We are interested in the formation of bound states through the emission of a vector $V^a$:
\beq {\rm DM}_i(P_1) + {\rm DM}_j(P_2) \to B_{i'j'} + V^a (K).\eeq
In the non-relativistic limit, we write the 4-momenta as 
\beq\label{eq:momenta}
P_1 \simeq (\MDM + \frac{p^2_1}{2\MDM}, \vec p_1),\qquad
P_2 \simeq (\MDM + \frac{p^2_2}{2\MDM}, \vec p_2),\qquad
K=(\omega,\vec k)\eeq
with $\omega= \sqrt{k^2 +M_V^2}$ where $M_V$ is the vector mass. 
In the center-of-mass frame $\vec p_2 = - \vec p_1$ and
the momentum of each DM particle is  $p = \MDM  v_{\rm rel}/2$.
Conservation of energy reads
\be
\frac {p^2}{\MDM} = \frac{k^2}{2(2\MDM -E_B)}- E_B +\omega
\label{eq:kinematics}
\ee
where $E_B= 2 \MDM - M_B>0$ is the binding energy.
The first term on the right-hand side is the recoil energy of the bound state that is negligible in what follows,
such that energy conservation approximates to
$\omega \approx E_B + \sfrac{\MDM v_{\rm rel}^2}{4}$.

\medskip

The diagrams in fig.~\ref{fig:bsf} contribute to the amplitude.
In the non-relativistic limit the first two diagrams describe the usual dipole approximation,
which gives a cross section for bound state formation proportional to $\alpha^5$,
times a sizeable Sommerfeld correction.
The third diagram is only present when the gauge interaction is non-abelian and was considered in~\cite{Slatyer} in the DM context.
We generalise their formul\ae{} to general non-abelian gauge theories, including
the regime where the initial velocity is not negligible as required for computing the thermal relic abundance.  
The diagrams of fig.~\ref{fig:bsf} generate the non-relativistic Hamiltonian~\cite{segno}
\be\label{eq:HI}
H_I=-\frac{g}{M_\chi} \Big( \vec A^a(x_1)\cdot \vec p_1 T^a_{i'i}\delta_{jj'}+ \vec A^a(x_2) \cdot \vec p_2 \overline{T}^{a}_{j'j}\delta_{ii'} \Big)+ \Big( g \alpha \vec{A}^a(0)\cdot \hat{r}\,e^{-M_a r}\Big)\;T^b_{i'i} \overline{T}^c_{j'j} f^{abc}
\ee
where $T$ and $\overline{T}$ are the generators in the representation of
particles 1 and 2 respectively; the indexes $a,b,c$ run over the vectors in the adjoint, and the indexes $i,j,i',j'$ over DM components.

In Born approximation we get the following cross section for the formation of a bound state with quantum numbers $n\ell m$:
\beq
\sigma^{n\ell m}_{{\rm bsf}} v_{\rm rel}=\sum_a(\sigma^{n\ell m}_{{\rm bsf}} v_{\rm rel})_a\eeq
where
\beq
(\sigma^{n\ell m}_{{\rm bsf}} v_{\rm rel})_a=\frac{2\alpha }{ \pi}\frac{k}{\MDM^2}\int d\Omega_k\sum_\sigma\left| 
\epsilon_\mu^a(k,\sigma) \mathscr{A} ^\mu_{ p, n\ell m}\right|^2
\label{eq:xsecgen}
\eeq
For massive gauge bosons the polarization vectors satisfy
\be
\sum_\sigma \epsilon^a_\mu\epsilon_\nu^{a*}=-\left(\eta_{\mu\nu}-\frac{K_\mu K_\nu}{M_a^2}\right).
\ee
The transition amplitude $\mathscr{A}^\mu$, computed from the matrix element of the interaction Hamiltonian,
satisfies $K_\mu\mathscr{A} ^\mu=0$ because of current conservation. 
Therefore the unpolarized cross section can be rewritten in terms of the spatial terms as 
\be\label{eq:sva}
(\sigma^{n\ell m}_{{\rm bsf}} v_{\rm rel})_a=\frac{2\alpha }{ \pi}\frac{k}{\MDM ^2}\int d\Omega_k\bigg(
| \vec{\mathscr{A}}_{ p, n\ell m}^a|^2 -\frac{\big| \vec k \cdot \vec{\mathscr{A}}_{ p, n\ell m}^a\big|^2}{k^2+M_a^2}\bigg).
\ee
In the dipole approximation,\footnote{The dipole approximation is valid if the wave-length of the photon is larger than the size of the bound state. 
As discussed in \cite{An:2016gad} the most relevant bound states are approximately
Coulomb-like so that the binding energy is $\alpha_{\rm eff}^2 \MDM/(4n^2)$ and the size the  Bohr radius $a_0=2 n^2/(\alpha_{\rm eff} \MDM)$. 
If follows that when the binding energy dominates over the initial kinetic energy the dipole approximation is always
satisfied. The dipole approximation fails for $v_{\rm rel}^2 \gg \alpha_{\rm eff}$. When this condition is verified the value of the cross-section is however small.} that will be used throughout, the spatial part of the transition matrix in the center-of-mass frame is
\beq
\vec{\mathscr{A}}_{p, n\ell m}^a=
\frac{1}{2}\left(T_{i'i}^a\delta_{jj'}-\overline{T}_{j'j}^{a}\delta_{ii'}\right)\vec{\cal J}^{ij,i'j'}_{p,n\ell m}-i
\left(T_{i'i}^b\overline{T}_{j'j}^cf^{abc}\right)\vec{\cal T}^{ij,i'j'}_{p,n\ell m}
\label{eq:genamp}
\eeq
where we have defined the overlap integrals between the initial state wave function $\phi_{p\ell,ij}(\vec r)$ 
and
the wave function $\psi_{n\ell m,i'j'}(\vec r)$ of the desired bound-state:
\begin{align}\label{eq:overlapint}
&\vec {\cal J}^{\,ij,i'j'}_{ p,n\ell m}\equiv \int d^3r\, \psi_{n\ell m,i'j'}^*  \vec \nabla\,\phi_{ p,ij}\\
&\vec {\cal T}^{\,ij,i'j'}_{ p,n\ell m}\equiv \frac{\alpha \MDM}{2}  \int d^3r\, \psi_{n\ell m,i'j'}^*\, \hat{r}\,e^{-M_ar}\,\phi_{ p,ij}\,.
\end{align}
The dipole approximation imposes the selection rule $\Delta L=1$. Since in the non-relativistic limit spin is also conserved 
this implies that $s$-wave bound states can only be produced from two DM particles in an  initial $p$-wave state.
Furthermore, $p$-wave bound states can be produced from $s$ and $d$-waves. With this in mind we get the following overlap integrals for the production of bound states in $s$-wave configuration:
\begin{subequations}
\begin{eqnarray}
\vec{\cal J}_{ p,n00}^{\,ij,i'j'}&=&-\frac{1}{\sqrt 3}\left(\int  r^2dr R_{p1,ij}(r)\partial_rR^*_{n0,j'i'}(r)\right)(\hat e_0+\hat e_++\hat e_-),\\
\vec{\cal T}_{ p,n00}^{\,ij,i'j'}&=&\frac{\alpha \MDM}{2\sqrt3} \left(\int  r^2dr R_{p1,ij}(r)e^{-M_a r}R^*_{n0,j'i'}(r)\right)(\hat e_0+\hat e_++\hat 
e_-)\,.
\end{eqnarray}
where $\hat e_0\equiv\hat z$ and $\hat e_\mp\equiv \pm\frac{1}{\sqrt2}(\hat x\mp i\hat y)$.
For production of bound states in a $p$-wave configuration starting from an $s$-wave one we get 
\begin{align}
&\vec{\cal J}^{\,ij,i'j'}_{\bold p,n1\pm1}=\frac{1}{\sqrt 3}\left(\int  r^2dr R_{p0,ij}(r)\partial_rR^*_{n1,j'i'}(r)\right)\hat e_{\mp},\\
&\vec{\cal J}^{\,ij,i'j'}_{\bold p,n10}=\frac{1}{\sqrt 3}\left(\int  r^2dr R_{p0,ij}(r)\partial_rR^*_{n1,j'i'}(r)\right)\hat e_{0},\\
&\vec{\cal T}^{\,ij,i'j'}_{\bold p,n1\pm1}=\frac{\alpha \MDM}{2\sqrt3} \left(\int  r^2dr R_{p0,ij}(r)e^{-M_a r}R^*_{n1,j'i'}(r)\right)\hat e_{\mp},\\
&\vec{\cal T}^{\,ij,i'j'}_{\bold p,n10}=\frac{\alpha \MDM}{2\sqrt3} \left(\int  r^2dr R_{p0,ij}(r)e^{-M_a r}R^*_{n1,j'i'}(r)\right)\hat e_0\,.
\end{align}
The amplitudes for producing a $p$-wave bound state starting from a $d$-wave configuration are
\begin{align}
&\vec{\cal J}^{\,ij,i'j'}_{\bold p,n1\pm1}=-\frac{1}{\sqrt5}\left[\int r^2dr R_{p2,ij}(r)\left(\partial_r-\frac{1}{r}\right)R^*_{n1,j'i'}(r)\right]\left(\sqrt2\hat e_\pm+\frac{\hat e_\mp}{\sqrt3}+\hat e_0\right),\\
&\vec{\cal J}^{\,ij,i'j'}_{\bold p,n10}=-\frac{1}{\sqrt5}\left[\int r^2dr R_{p2,ij}(r)\left(\partial_r-\frac{1}{r}\right)R^*_{n1,j'i'}(r)\right]\left(\hat e_++\hat e_-+\frac{2}{\sqrt3}\hat e_0\right),\\
&\vec{\cal T}^{\,ij,i'j'}_{\bold p,n1\pm1}=\frac{ \alpha \MDM }{2\sqrt5}\left[\int r^2dr R_{p2,ij}(r)e^{-M_a r}R^*_{n1,j'i'}(r)\right]\left(\sqrt2\hat e_\pm+\frac{\hat e_\mp}{\sqrt3}+\hat e_0\right),\\
&\vec{\cal T}^{\,ij,i'j'}_{\bold p,n10}=\frac{\alpha \MDM }{2\sqrt5}\left[\int r^2dr R_{p2,ij}(r)e^{-M_a r}R^*_{n1,j'i'}(r)\right]\left(\hat e_++\hat e_-+\frac{2}{\sqrt3}\hat e_0\right)\,.
\end{align}
\end{subequations}
Plugging these amplitudes in eq.~\eqref{eq:sva}, performing the angular integral, averaging over initial states and summing over final states we get the cross sections for the formation of $s$-wave bound states:  
\be
\begin{aligned}\label{eq:n0}
(\sigma^{n0}_{{\rm bsf}} v_{\rm rel})_a^{p\to s}&=\frac{8}{3}\frac{\alpha k}{\MDM^2}\left(1-\frac{k^2}{3\omega^2}\right)\times \\
&\times \Bigg|\int r^2drR_{p1,ij}\Bigg(\frac{1}{2}\left(T_{i'i}^a\delta_{jj'}-\overline{T}_{j'j}^{a}\delta_{ii'}\right) \partial_r
+i\,\frac{\alpha\MDM}{2} \left(T_{i'i}^b \overline{T}_{j'j}^cf^{abc}\right) e^{-M_ar}\Bigg)R^*_{n0,j'i'}\Bigg|^2
\end{aligned}
\ee
For $p$-wave bound states we get
\begin{subequations}\label{eq:n1}
\be
(\sigma^{n1}_{{\rm bsf}} v_{\rm rel})_a=(\sigma^{n1}_{{\rm bsf}} v_{\rm rel})_a^{s\to p}+(\sigma^{n1}_{{\rm bsf}} v_{\rm rel})_a^{d\to p}
\ee
where $(\sigma^{n1}_{{\rm bsf}} v_{\rm rel})_a^{s\to p}$ and $(\sigma^{n1}_{{\rm bsf}} v_{\rm rel})_a^{s\to p}$ are the cross sections from initial states in $s$ and $d$-wave respectively.
Their explicit values are 
\be
\begin{aligned}
(\sigma^{n1}_{{\rm bsf}} v_{\rm rel})_a^{s\to p}&=8\frac{\alpha k}{\MDM^2}\left(1-\frac{k^2}{3\omega^2}\right)\Bigg|\int r^2drR^*_{n1,j'i'} \\
&\times\Bigg(\frac{1}{2}\left(T_{i'i}^a\delta_{jj'}-\overline{T}_{j'j}^{a*}\delta_{ii'}\right) \partial_r
-i\, \frac{\alpha\MDM}{2} \left(T_{i'i}^b\overline{T}_{j'j}^cf^{abc}\right) e^{-M_ar}\Bigg)R_{p0,ij}\Bigg|^2
\end{aligned}
\ee
\be
\begin{aligned}
(\sigma^{n1}_{{\rm bsf}} v_{\rm rel})_a^{d\to p}&=\frac{16}{5}\frac{\alpha k}{\MDM^2}\left(1-\frac{k^2}{3\omega^2}\right) 
 \Bigg|\int r^2drR_{p2,ij}\times\\
&\times\Bigg(\frac{1}{2}\left(T_{i'i}^a\delta_{jj'}-\overline{T}_{j'j}^{a{*}}\delta_{ii'}\right)\left(\partial_r-\frac{1}{r}\right)
+i\,\frac{\alpha\MDM}{2} \left(T_{i'i}^b\overline{T}_{j'j}^cf^{abc}\right) e^{-M_ar}\Bigg)R^*_{n1,j'i'}\Bigg|^2
\end{aligned}
\ee
\end{subequations}
If DM are scalars,  the wave function is symmetric under exchange of identical scalars.
Real (complex) scalars have $g_\chi = d_R$ ($2d_R$) degrees of freedom.
Bound states of scalars have $S=0$.
For $s$ ($p$)-wave bound states this implies that the gauge part of the wave function is symmetric (anti-symmetric).
The cross-sections for  bound state formation are again given by eq.~(\ref{eq:n1}).

\subsection{Group algebra}\label{sec:massless}
Assuming that the global group $G$ is unbroken
(such that vectors are either massless or have a common mass), group algebra allows to simplify the above formul\ae{}.
We assume that DM is a particle $\chi_i$ in the representation $R$ of $G$, labeled by an index $i$, and
we focus on $\chi_i \overline\chi_j$ bound states so that $\overline{T}^a=-T^{a*}$.
Both the initial state and each bound state can be decomposed
into irreducible representations of $G$,
times the remaining spin and spatial part.
So the two-body DM states $\chi_i \overline\chi_j$ fill the representations $J$ contained in $R\otimes \overline R = \sum_J J$.
Each representation $J$ is labeled by an index $M$.
The change of basis is described the the coefficients $\hbox{CG}^M_{ij} \equiv \langle J , M |R,i ; \overline R , j\rangle$ of the group $G$.
For $G=\SU(2)_L$ these are the  Clebsh-Gordon coefficients usually written as $\langle j,m | j_1, m_1;j_2,m_2\rangle$.
For the singlet representation one has 
$\hbox{CG}_{ij}=\sfrac{\delta_{ij}}{\sqrt d_R}$
and for the adjoint representation one has
$\hbox{CG}^a_{ij}=\sfrac{T^a_{ij}}{\sqrt T_R} $.
In the new basis, where $ij$ is replaced by $M$ and $i'j'$ by $M'$, the bound-state formation amplitudes of
eq.~\eqref{eq:genamp} becomes
\beq
\vec{\mathscr{A}}_{p, n\ell m}^{aMM'}=
C_{\cal J}^{aMM'}\vec{\cal J}_{p,n\ell m}+
C_{\cal T}^{aMM'}\vec{\cal T}_{p,n\ell m}
\eeq
where the group-theory part has been factored out in the coefficients
\begin{eqnsystem}{sys:CJT}
C_{\cal J}^{aMM'} &\equiv &\frac{1}{2}\,\hbox{CG}^M_{ij}\hbox{CG}^{M'*}_{i'j'}
 (T_{i'i}^a\delta_{jj'}+T_{j'j}^{a*}\delta_{ii'})
 =\frac{1}{2} \Tr[\hbox{CG}^{M'}\{\hbox{CG}^M,T^a\}]
 \\
C_{\cal T}^{aMM'} &\equiv& i\,\hbox{CG}^M_{ij}\hbox{CG}^{M'*}_{i'j'}  (T_{i'i}^bT_{jj'}^cf^{abc})=
i\Tr\Big[\hbox{CG}^{M'}\, T^b\,\hbox{CG}^M\,T^c\Big]f^{abc} \label{eq:nonabel}
\end{eqnsystem}
that holds separately for each initial channel $J$ and  final channel $J'$. 
In many cases of interest the two tensors are proportional to each other.
The overlap integrals $\cal{J}$, $\cal{T}$ are the same of eq.~\eqref{eq:overlapint}, 
but now containing only the spatial part of the wave functions. With these notations the cross sections of eq.~(\ref{eq:n0}) and (\ref{eq:n1}), in a given channel $(J,M)\to (J',M')$, become 
\begin{eqnsystem}{sys:sigmasgen}
&(\sigma^{n0}_{{\rm bsf}} v_{\rm rel})_{aMM'}^{p\to s}=\frac 8 3\frac{\alpha k}{\MDM^2}\left(1-\frac{k^2}{3\omega^2}\right)\left|\int r^2dr R_{p1}\left(C_{\cal J}^{aMM'}\partial_r-C_{\cal T}^{aMM'}\frac{\alpha\MDM}{2}e^{-M_a r}\right)R^*_{n0}\right|^2 
\\
&(\sigma^{n1}_{{\rm bsf}} v_{\rm rel})_{aMM'}^{s\to p}=8\frac{\alpha k}{\MDM^2}\left(1-\frac{k^2}{3\omega^2}\right)\left|\int r^2drR^*_{n1}\left(C_{\cal J}^{aMM'} \partial_r+C_{\cal T}^{aMM'}\frac{\alpha\MDM}{2}  e^{-M_ar}\right)R_{p0}\right|^2
\\
&(\sigma^{n1}_{{\rm bsf}} v_{\rm rel})_{aMM'}^{d\to p}=\frac{16}{5}\frac{\alpha k}{\MDM^2}\left(1-\frac{k^2}{3\omega^2}\right)\left|\int r^2drR_{p2}\left(C_{\cal J}^{aMM'}\left(\partial_r-\frac{1}{r}\right)-C_{\cal T}^{aMM'}\frac{\alpha\MDM }{2} e^{-M_ar}\right)R^*_{n1}\right|^2.~~~~~
\end{eqnsystem}
In the special case $1\to \hbox{adj}$ (namely,  the initial state is a gauge singlet, such that the bound state is an adjoint)
the group theory factors are proportional to each other,
$C_{\cal J}^{a M M'} \propto C_{\cal T}^{a M M'}$,
so that the inclusive cross-section remains a perfect square:
\begin{equation}\label{eq:bsfC10}
 \sum_{a M M'} \left|  C_{\cal J}^{a M M'}  + \gamma C_{\cal T}^{a M M'} \right|^2=  \frac {T_R d_{{\rm adj}}}{d_R}\left|1\mp\frac{ \gamma}2  T_{\rm adj}  \right|^2.
\end{equation}
The $+$ sign corresponds to the opposite $ \hbox{adj}\to 1$ process.
The same simplification holds for any $\SU(2)_L$ representation, 
because in the product of two $\SU(2)_L$ representations each irreducible representation appears only once.
The relevant $\SU(2)_L$ group factors are listed in table~\ref{tab:C5}.
Furthermore, the simplification also holds for the $\SU(3)_c$ representations that we will encounter later,
and the  relevant group $\SU(3)_c$ factors are listed in table~\ref{tab:Ccolor}.


\subsection{Massless vectors}\label{bsfmassless}
The overlap integrals in the spatial part of the amplitudes for bound state formation can be analytically computed if
vectors are massless.

The initial states are assumed to be asymptotically plane-waves with momentum $\vec p$, distorted by the potential in channel $J$
where $\alpha_{\rm eff}=\lambda_i \alpha$ 
with $\lambda_i = \lambda_J$ given by eq.~\eqref{eq:VQ}. 
The initial state wave function in a Coulomb-like potential  is given in eq.\eq{FreeCoulomb}.

The final states are assumed to be bound states in channel $J'$ 
in a Coulomb potential with  $\alpha_{\rm eff}=\lambda_f \alpha$ 
and $\lambda_f = \lambda_{J'}$. 
We use a basis of eigenstates of angular momentum, parameterized by the usual $\ell,m$ indeces.
The  bound state wave functions are given in  eq.~(\ref{eq:psiCoulomb}), and are
analytic continuations of the free-state wave functions.

Plugging these wave functions into the overlap integrals
we get the cross section for the production of the various bound states. 
We are interested in the cross-section averaged over initial states 
and summed over final gauge bosons and bound states components.
For the lowest lying bound state with $n=1$, $\ell=0$ and spin $S$ we get
\beq\label{eq:bsf10}
(\sigma v_{\rm rel})_\text{bsf}^{n=1,\ell=0} =   \sigma _0   \lambda_i (\lambda_f \zeta)^5 \frac{2 S +1}{g_\chi^2} \, \frac{2^{11} \pi (1+\zeta ^2 \lambda_i^2) e^{-4 \zeta  \lambda_i  \text{arccot}(\zeta  \lambda_f )}}{3 (1+\zeta ^2 \lambda_f^2)^3
   \left(1-e^{-2 \pi  \zeta  \lambda_i}\right)}\times \sum_{a M M'} \left|  C_{\cal J}^{a M M'}  + \frac 1{\lambda_f} C_{\cal T}^{a M M'} \right|^2
\eeq
where $\sigma_0=\pi \alpha^2/\MDM^2$ and $\zeta=\alpha/v_{\rm rel}$.
For the bound states with $n=2$ and $\ell=\{0,1\}$ we get
\begin{align}
\label{eq:bsfI0001}
&  (\sigma v_{\rm rel})_\text{bsf}^{n=2, \ell=0} = 
 \sigma _0 \lambda_i   \lambda_f^5
 \frac{2 S +1}{g_\chi^2} \, \frac{2^{14} \pi  \zeta ^5  \left(\zeta ^2 \lambda_i^2+1\right) e^{-4 \zeta 
   \lambda_i \text{arccot}(\sfrac{\zeta  \lambda_f}{2})} }{3 \left(\zeta ^2
   \lambda_f^2+4\right)^5 \left(1-e^{-2 \pi  \zeta  \lambda_i}\right) }  \\ \nonumber
& \times \sum_{aMM'}   \left|C_{\cal J}^{a M M'}  \left(\zeta ^2
   \lambda_f \left(\lambda_f-2 \lambda_i\right)-4\right)+C_{\cal T}^{a M M'} \left(\zeta ^2   \left(3 \lambda_f-4 \lambda_i\right)-\frac{4}{\lambda_f}\right)   \right|^2 \,, \\  \nonumber
& (\sigma v_{\rm rel})_\text{bsf}^{n=2, \ell=1}  =  \sigma_0\lambda_i\lambda_f^5  \frac{2 S +1}{g_\chi^2}\frac{2^{12} \pi \alpha \zeta ^7 
   e^{-4 \zeta  \lambda_i \text{arccot}(\zeta\lambda_f/2)}}{9 \left(\zeta ^2\lambda_f^2+4\right)^5 \left(1-e^{-2 \pi  \zeta \lambda_i}\right)} \times \nonumber \\ 
   &\times \sum_{aMM'}   \bigg[   \bigg|C_{\cal J}^{a M M'}  \bigg(\lambda_f (\zeta ^2 \lambda_i (3
   \lambda_f-4 \lambda_i)+8) \bigg. 
 \bigg. -12\lambda_i\bigg)  +   C_{\cal T}^{a M M'} \left(\zeta ^2 (-3 \lambda_f^2+12 
   \lambda_f   \lambda_i-8 \lambda_i^2)+4\right)\bigg|^2 + \nonumber \\ 
  &+ 2^5 (\zeta ^2 \lambda_i^2+1) (\zeta ^2 \lambda_i^2+4)
   \bigg|C_{\cal J}^{a M M'} \lambda_f+2  C_{\cal T}^{a M M'}\bigg|^2   \bigg] .
\end{align}
In the last equation we have separated the contribution of the $s$-wave and $d$-wave initial state.
These formulas apply both for Dirac and Majorana particles,
and in all cases relevant for us the sums can be performed as summarized in tables~\ref{tab:C5} and~\ref{tab:Ccolor}.

\smallskip

In the limit $\lambda_i=0$ where the Sommerfeld correction is ignored,
the cross section for producing a bound state with $\ell=0$ is
of order $\alpha^2/\MDM^2$ times a $(v_{\rm rel}/\alf)^2$ suppression at $v_{\rm rel}\ll \alf$ as expected for production from a $p-$wave;
the cross section for producing a bound state with $\ell=1$ does not have this suppression for $C_{\cal T}\neq 0$.

The formul\ae{} above simplify in the limit of large and small velocities. For the ground state one finds
\begin{equation}\label{eq:bsf10exp}
(\sigma v_{\rm rel})_\text{bsf}^{n=1,\ell=0} =  \sigma _0
\frac{2 S +1}{g_\chi^2} \, \frac{2^{11} \pi }{3 } \sum_{a M M'} \left|  C_{\cal J}^{a M M'}  + \frac 1{\lambda_f} C_{\cal T}^{a M M'} \right|^2\times
\left\{\begin{array}{ll}
\displaystyle{ \frac {\lambda_i^3\alpha}{\lambda_fv_{\rm rel}}e^{-4\lambda_i/\lambda_f}} & v_{\rm rel}\ll \lambda_{i,f} \alpha\\
\displaystyle{\frac{\lambda_f^5 \alpha^4}{2 \pi v_{\rm rel}^4}} & v_{\rm rel}\gg\lambda_{i,f} \alpha
\end{array}\right. 
\end{equation}
For large velocities the cross-section is proportional $\alpha \alf^5  /v_{\rm rel}^4$. 

\subsection{Approximate formul\ae{} for massive vectors}\label{sec:massive}
The cross sections for producing bound states in a Yukawa potential can
be obtained by computing numerically the wave functions (or using the wave functions in Hulthen approximation, listed in the appendix),
and by computing numerically the overlap integrals.
As this is somehow cumbersome, we discuss how massless formul\ae{} can be readapted, with minor modifications,
to take into account the main effects of vector masses. 
We start considering the case where the vectors have a common mass $M_V$ and the group theory structure is identical to the massless case.

The initial state wave function remains approximately Coulombian as long as  $M_V\ll \MDM v_{\rm rel}$.
Physically, this means that  the range of the force $1/M_V$ is much larger than the de Broglie wave-length of Dark Matter $\lambda^{-1}= \MDM v_{\rm rel}$.
One indeed can check that in this limit the Sommerfeld factor in eq.\eq{Som} is well approximated by its Coulombian limit $M_V=0$.
At finite temperature $T \sim \MDM v_{\rm rel}^2$, so that the Coulombian approximation holds for 
temperatures $T \gg \sfrac{M_V^2}{\MDM} $
which can be much lower than $M_V$. When this condition is violated, the modification of the shape of the potential leads to a scaling of the cross section 
with velocity as $v_\text{rel}^{2\ell}$, where $\ell$ is the angular momentum of the initial state wave function.  
Thus, for the $1s$ bound state, which is created from a $p$ wave state,  the scaling is $v_\text{rel}^{2}$.  
Therefore, the cross section is velocity suppressed and small after thermal average at late times. 
On the other hand $p$-wave bound states which are formed from an $s$-wave initial state approach a constant value.

\smallskip

Next, we consider bound states.  Eq.\eq{EBHul} shows that bound states are well approximated
by the Coulombian $M_V=0$ limit if $M_V \ll \MDM\alf$.
This condition can be alternatively obtained from the analogous condition for free states by replacing $v_{\rm rel}\to\alf$,
since this is the typical velocity in a bound state. 
In the limit of small $M_V \ll \alf \MDM$ all binding energies undergo a small common shift $-\alf M_V$ as discussed around eq.\eq{EBsmallMV}. 

\medskip

In summary for $T \gg \sfrac{M_V^2}{\MDM} $ the main effect of vector masses is the kinematical suppression of the cross section for bound-state formation,
which blocks the process if $M_V$ is bigger than the total accessible energy.
This effect  is approximately captured by
\begin{equation}\label{eq:kine}
\frac{\sigma  (\DM\DM \to B V)}{  \sigma (\DM\DM \to B V)|_{M_V=0} }\approx \frac 3 2  \frac{ k}{\omega} \left(1- \frac {k^2}{3 \omega^2}\right)
\qquad\text{  for  } \qquad x < \frac{\MDM^2}{M_V^2}
\end{equation}
where $K_\mu = (\omega,\vec k)$ is the massive vector  quadri-momentum as in eq.\eq{momenta}.
The  parenthesis take into account the emission of the third polarization of a massive vector. 
The bound state formation gets suppressed or blocked when $\omega$ becomes of order  $M_V$.

In our applications we will need the cross sections below the critical temperature at which $\SU(2)_L$ gets broken.
In this case the masses are not degenerate: one has   $M_W\approx M_Z$ and $M_\gamma=0$.
It becomes important to include emission of photons and eq.~\eqref{eq:kine} becomes
\begin{equation}
\frac{\sigma  (\DM\DM \to B V)}{  \sigma (\DM\DM \to B V)|_{M_V=0} }\approx  \frac{ k}{\omega} \left(1 - \frac {k^2}{3 \omega^2}\right) \left(1+\frac{\cos^2 \theta_{\rm W}}{2}\right)+\frac{\sin^2 \theta_{\rm W}}{3} \, ,
\end{equation}
where the first term takes into account the emission of the $W$ and $Z$ bosons while the last term corresponds to the photon emission. 

One extra effect is that the charged components of the DM electroweak multiplet get split from the
neutral component and become unstable. In the cases of interest discussed later, 
the resulting decay width negligibly affects the cosmological relic DM density.

\section{Annihilations of DM in bound states, and their decays}\label{sec:decays}
The two DM particles bound  in a potential $V=-\alf e^{-M_V r}/r$ can annihilate to SM particles, such that the bound state decays.
We will refer to this process as `annihilation' rather than `decay'.
Analogously to  quarkonium in QCD, the rate  is
\beq \Gamma_{\rm ann} \sim \alf^3 \alpha^2_{\rm SM} \MDM \circa{>} 10^{-8}\MDM.\eeq
This is typically much faster than the Hubble rate 
\beq 
H  = \sqrt{\frac{4 \pi^3 g_{\rm SM}}{45} }\frac{T^2}{M_{\rm Pl}} \approx  2 ~10^{-18} \MDM \times \frac{\MDM}{\TeV} \qquad
\hbox{at }T\approx \frac{\MDM}{25}.\eeq
Nevertheless breaking of bound states in the thermal plasma can have a rate $\Gamma_{\rm break}(T)$
which is  as fast as $\Gamma_{\rm ann}$ at the freeze-out temperature.
So we need to compute the annihilation rates in order to obtain the branching ratios in eq.\eq{B1}.
We assume that DM is heavy enough that we can ignore the masses of SM particles produced in annihilations of DM bound states.

The group-theory factors are analogous to the one encountered in section~\ref{Sommerfeld}
when computing Sommerfeld-enhanced DM annihilations to SM particles.
As already discussed, the DM-DM bound states $\chi_i \overline\chi_j$ fill the representations $J$ contained in
$R\otimes\overline R = \sum_J J $, and the bound state $B_M$ in representation $J$ with index $M$ 
is given by $\hbox{CG}_{ij}^M \chi_i \overline\chi_j$.

\subsection{Annihilations of spin 0 bound states with  $\ell=0$}
We assume that the gauge group is unbroken and that DM is much heavier than SM particles.
The annihilation rate of a spin-0 bound state $B_{n\ell}^M$ with $\ell=0$
into two vectors $V^a V^b$, summed over all their
components $a,b$ is
\begin{equation}\label{eq:Gamman0S0}
\Gamma_{\rm ann} = \Gamma(B^M_{n0}\to VV)=\alpha^2\frac {|R_{n0}(0)|^2}{F^2 \MDM^2} \sum_{a,b} {\rm Tr}\left[\hbox{CG}^M
 \frac {\{T^a_R,T^b_R\}}2\right]^2
\end{equation}
where $T^a_R$ is the generator in the DM representation $R$, and
$R_{n\ell}(r)$ is the radial wave function of the bound state normalized as $\int_0^\infty |R_{n\ell}(r)|^2 r^2 dr=1$;
$F=1(2)$ for distinguishable (identical) DM particles. For Majorana particles the amplitude is 1/2 the one of Dirac particles while the wave function at the origin is $\sqrt{2}$ so that the total rate is 1/2 the one of Dirac particles.

In general $R\otimes\overline R $ always contains the singlet and the adjoint representation,
so we evaluate explicitly the group-theory factors that determine the annihilation rates of these specific bound states.
\begin{itemize}
\item For a gauge-singlet bound state one has ${\rm CG}_{ij}= \delta_{ij}/\sqrt{d_R}$ such that
its annihilation rate is
\begin{equation}
\Gamma(B_{n0}\to VV)= \alpha^2\frac {|R_{n0}(0)|^2}{F^2 \MDM^2} \frac {T_R^2 d_{\rm adj}}{d_R} 
\end{equation}
where $\Tr T^a_R T^b_R = T_R \delta^{ab}$.

\item For a bound state $B^a$ in the adjoint representation of $G$ one finds
\begin{equation}\label{eq:gamma0dec}
\Gamma(B^a_{n0}\to VV)= \alpha^2\frac {|R_{n0}(0)|^2}{16 F^2\MDM^2} \frac {\sum_{abc} d_{abc}^2}{ d_{\rm adj}} 
\end{equation}
where $d_{abc}=2 {\rm Tr}[{\rm CG}^a \{T^b, T^c\}]$. This is zero if $G=\SU(2)$.
Indeed the triplet bound state for $\SU(2)$ has spin-1 and cannot decay into massless vectors. 
\end{itemize}
The annihilation rate into scalars is given by one half of the above expression.

\smallskip

The previous formulas hold for a generic Yukawa potential.
In the Coulomb limit the wave functions can be explicitly evaluated, obtaining
\begin{equation}
\frac {|R_{n0}(0)|^2}{\MDM^2}=F\,\frac{\MDM \alpha _{\text{eff}}^3}{2 n^3}.
\end{equation}
Approximating the Yukawa potential with the  Hulthen potential one finds
\begin{equation}
\frac {|R_{n0}(0)|^2}{\MDM^2}=F\,\frac{\MDM \alpha _{\text{eff}}^3}{2 n^3}
\left(1-\frac{\kappa ^2 n^4 M_V^2}{\MDM^2 \alpha _{\text{eff}}^2}\right).
\end{equation}

\subsection{Annihilations of spin 1 bound states}
In view of the Landau-Yang theorem, spin-1 bound states cannot annihilate into $VV$.
They can annihilate into pairs of SM fermions and scalars (or equivalently longitudinal gauge bosons). For fermions
\begin{equation}
 \Gamma(B^M_{n0}\to f_i \overline{f_j})=\frac{\alpha^2}{6}\frac {|R_{n0}(0)|^2}{F^2 \MDM^2} \sum_{a} |{\rm Tr}\left[\hbox{CG}^M
 T^a_R \right] T _{{\rm SM}ij}^a|^2
\end{equation}
where $T_{\rm SM}^a$ are the gauge  generators of the considered SM fermion. 
The rate is different from zero only for bound state in the adjoint representation
($\hbox{CG}^a_{ij}= T^a_{ij}/\sqrt{T_R}$). Summing over the components of $f$ we get
\begin{equation}
 \Gamma(B^a_{n0}\to f \overline{f})=\frac{\alpha^2}{6}\frac {|R_{n0}(0)|^2}{ F^2 \MDM^2}T_R T_{\rm SM}
\end{equation}
that should be multiplied by the multiplicity of final states: the SM contains $ 3(3+1)$ fermionic $\SU(2)_L$ doublets.
If DM has hypercharge, the annihilation rate receives the extra contribution 
\begin{equation}
\Delta\Gamma(B^a_{n0}\to f \overline{f})=   \frac {\alpha_Y^2} 6 \frac {|R_{n0}(0)|^2}{F^2 \MDM^2}d_R Y_Q Y_f.
\end{equation}
Spin-1 singlet resonances can also decay into three vectors, but with a suppressed rate
\beq 
\Gamma(B_{n0}\to VVV)  =\frac {\sum_{abc} d_{abc}^2}{36 d_R} \frac{\pi^2-9}{\pi} \alpha^3\frac {|R_{n0}(0)|^2}{F^2 \MDM^2}.
 \eeq

\subsection{Annihilations of bound states with $\ell>0$}
The annihilation rate of bound states with orbital angular momentum $\ell>0$
is suppressed by higher powers of $\alpha$.
For example spin-1 bound states annihilate into vectors as
\begin{equation}
\Gamma(B^M_{n 1}\to VV)=9 \alpha^2\frac {|R_{n1}'(0)|^2}{F^2\MDM^4}\frac 1 {d_B} \sum_{a,b} {\rm Tr}\left[\hbox{CG}^M \frac {\{T^a,T^b\}}2\right]^2
\end{equation}
where in the massless limit the derivative of the wave function at the origin contains the suppression factor
\begin{equation}
\frac {|R_{21}'(0)|^2}{\MDM^4}=F\,\frac {\alpha_{\rm eff}^5}{24}{\MDM} 
\end{equation}
Annihilations of spin-0 bound states with $\ell=1$ into fermions and scalars are similarly suppressed.
A greater suppression applies to bound states with $\ell>1$.
We will not need to compute these suppressed annihilation rates because states with $\ell>0$ 
undergo faster decays into lower bound states, as discussed in the next section.

\subsection{Decays of  bound states}
We next consider decays of a DM bound state into another lighter bound state.
This is analogous to decays of excited state of the hydrogen atom.\footnote{With the important
difference that Dark Matter (unlike hydrogen at recombination)
has a small number density at freeze-out, such that vectors emitted at bound state formation (unlike photons) or from bound states
have a negligible impact on the plasma.}

The decay rate of a $2s$ state into the corresponding $1s$ state is suppressed, and negligible with respect to
its annihilation rate.

The decay rate of a $2p$ state into the corresponding $1s$ state is unsuppressed, and dominant with respect to
its annihilation rate.
The formula for the decay rate is related 
to the cross-section for bound state formation~\cite{Slatyer}:
the only difference is that the initial state is  not 
a free state, but a bound states with wave functions normalized to 1.
Explicitly
\begin{equation}
\Gamma(B^M_{21}\to B^{M'}_{10} +  V^a)= \frac{16}{9}\frac{\alpha k}{\MDM^2}\left|\int r^2dr R_{21}\left(C_{\cal J}^{aMM'}\partial_r-C_{\cal T}^{aMM'}\frac{\alpha\MDM}{2}e^{-M_a r}\right)R^*_{10}\right|^2.
\end{equation}
If $G=\SU(2)_L$ and at temperatures below the scale of electroweak symmetry breaking
the released binding energy is usually not enough
to emit a massive $\SU(2)_L$ vector $W$ or $Z$,
and only the photon can be emitted.  
\begin{equation}
\Gamma(2p\to 1s+  \gamma)= \alpha_{\rm em} \alpha_2^4 \MDM \left(\lambda_f^2- \frac {\lambda_i^2} 4\right) 
     \frac {512 \lambda_i^5 \lambda_f^{5}}{3(\lambda_i+2 \lambda_f)^8} \times \frac 1{3 d_B}\sum_{aMM'} \left|  C_{\cal J}^{a M M'}  + \frac{ C_{\cal T}^{a M M'}}{\lambda_f} \right|^2.
\end{equation}
having assumed that the bound state is well approximated by its Coulombian limit.

\section{Thermal effects}\label{Thermal}
So far we allowed for generic vectors mass.
The motivation is that all vectors acquire non-relativistic `thermal masses' in the early universe at finite temperature.
In the non-relativistic limit we are interested in electric potentials, and 
the relevant masses are the Debye masses, given by
\beq \label{eq:mT}
m_{{\rm U(1)}}^2 = \frac{11}{6}g_Y^2 T^2,\qquad 
m_{\SU(2)}^2 = \frac{11}{6}g_2^2T^2,\qquad 
m_{\rm SU(3)}^2 = 2 g_3^2 T^2. \eeq
This means that an attractive potential with $\alf = \lambda \alpha $
supports bound states with quantum number $n=1,2,\ldots$ if
\beq\label{eq:lambdaBS}
\lambda \ge \frac{T}{\MDM/25} n^2
 \left\{\begin{array}{ll}
1.7& \hbox{for $\SU(2)_L$}\\
1.0 & \hbox{for $\SU(3)_c$}
\end{array}\right.  .\eeq
Furthermore,
the $W^\pm$ and the $Z$  acquire mass 
from the electro-weak symmetry breaking.
Combining $\SU(2)_L$-breaking masses with thermal masses gives
a thermal mixing between $\gamma$ and $Z$.
At finite temperature the $\SU(2)_L$-breaking Higgs vev $v$
decreases until $\SU(2)_L$ is restored via a cross-over at $T>T_{\rm cr} \approx 155\GeV$.
This effect can be roughly approximated as
\beq v(T) = v \,{\rm Re} (1 - T^2/T_{\rm cr}^2)^{1/2} .\eeq
In reality, thermal corrections are a much more subtle issue.
We need to reconsider if/how the above naive approach applies at finite temperature.

\subsection{Sommerfeld enhancement at finite temperature}
Evolution of the DM states is affected by the presence of the thermal plasma.
At leading order in the couplings to a plasma one gets refraction
(in the case of the thermal plasma, this corresponds to thermal masses).
At second order one gets interactions with rates $\Gamma$
which exchange energy and other quantum numbers with the plasma,
and break quantum coherence among different DM components.
Thereby DM forms an open quantum system, 
which is not described by a wave function, but by a density matrix $\rho$.
Its evolution equation has the form
\beq \dot \rho =-i [H,\rho]+ \sum_L \Gamma_L (L \rho L^\dagger - \frac12 \{\rho,L^\dagger L\})
\label{eq:densitymatrix}
\eeq
where $L$ are Lindblad operators that describe the various interactions $\Gamma_L$~\cite{Lindblad}.
A gauge interaction with the plasma typically gives $\Gamma_L \sim \alpha^2 T^3/\MDM^2$.
Let us discuss breaking of quantum coherencies in the cases of interest.

\begin{figure}[t]
\begin{center}
\includegraphics[width=0.5\textwidth]{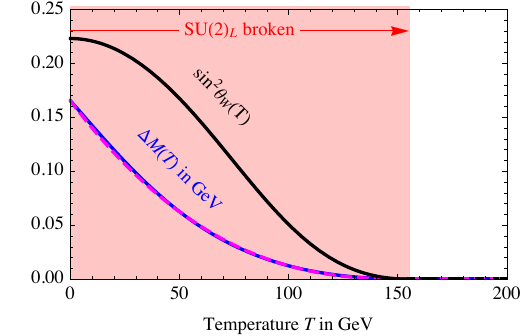}
\caption{\label{fig:DMT}\em  DM mass splitting (blue) and weak angle (black) at finite temperature.}
\end{center}
\end{figure}

In the $\SU(3)_c$ case, the Lindblad operators are proportional to the unit matrix in each 2-body sub-system with given quantum numbers.
Thereby coherencies within each sector with given total color
 is preserved, while contributions from different sectors to the total cross 
section must be summed incoherently.

In the $\SU(2)_L$ case, its breaking leads to loss of coherence within the components of a given representation.
For example, if DM is a $\SU(2)_L$ triplet with components $\chi_0$ and $\chi_\pm$,
a $\chi_0 \chi_0$ state can become $\chi_0\chi_+$ by interacting with soft $W^\pm$ vectors in the plasma.
From the point of view of exactly conserved quantum numbers, such as electric charge,
these are different sectors.
Thereby one has something intermediate between exact $\SU(2)_L$
(full coherence within each sector with given weak representation)
and badly broken $\SU(2)_L$ (coherence only between state with same electric charges).
An effect of this type is induced by the mass splitting among $\chi_0$ and $\chi_\pm$,
which randomises their relative phase. In a static situation this is equivalent to  loss of quantum coherence~\cite{hep-ph/9802387}.

So we compute the thermal contribution to the mass splitting between different components of $\SU(2)_L$ multiplets,
which was neglected in previous studies.
A fermion with mass $\MDM\gg M_V,T$ 
receives the following thermal correction to its mass, at leading order in $g$:
\beq \Delta M_T =\frac{ g^2 C}{4\pi^2} \frac{1}{\MDM} \int_0^\infty dk~k^2  \frac{ 2k^2+3M_V^2}{(k^2+M_V^2)^{3/2}} n_B(\sqrt{k^2+M_V^2}),\qquad n_B(E) = \frac{1}{e^{E/T}-1}.
\eeq
This correction is suppressed by $\MDM$ and  can be neglected for our cases of interest.
A correction not suppressed by $\MDM$ arises at higher order in $g$~\cite{0906.3052,LaineSommerfeld}, and
can be taken into account as follows.
In the limit $\MDM\gg M_V$ 
the one-loop quantum correction to the mass of a charged particle,
as computed from Feynman diagrams,
reduces to the classical Coulomb energy $U$ stored in the electric fields. 
For a single vector $A_\mu$  it is
\beq 
U = \int dV\bigg[\frac{(\nabla A_0)^2}{2} + \frac{M_V^2}{2}A_0^2\bigg]
= \frac{g^2}{8\pi} M_V + {\rm divergent}\qquad\hbox{where}\qquad
A_0 = \frac{g}{4\pi}\frac{e^{-M_V r}}{r}.\eeq
After summing over all SM vectors,
the mass difference between two DM components $i$ and $j$ with electric charges $Q_i$ and $Q_j$ in a generic 
Minimal Dark Matter model is~\cite{Cirelli:2007xd}
\beq \Delta M_{ij} =\frac{\alpha_2 }{2}[(Q_i^2-Q_j^2)s_{\rm W}^2(M_{Z}-M_\gamma)+(Q_i-Q_j)(Q_i+Q_j-2Y)(M_W-M_Z)].\eeq
The higher order thermal contribution is obtained by simply replacing $M_{\gamma}, M_Z,M_W$ and $s_{\rm W}$ with their thermal expressions.
For $Q_i=1$, $Q_j=Y=0$ the mass difference is plotted in fig.~\ref{fig:DMT} and well approximated by
\beq \Delta M(T) = 165\MeV \,\hbox{Re} (1-T/T_{\rm cr})^{5/2}.\eeq

\subsection{Bound-state formation at finite temperature}\label{finiteTbsf}

If thermal masses were naive masses, they could kinematically block bound-state  formation
$\chi\bar\chi \to B V$, when $M_V\sim gT $ is  bigger than the binding energy $E_B \sim \alpha^2 \MDM$.


However thermal masses are not naive masses.
Heuristically, one expects that a plasma cannot block the production of a
vector with  wave-length shorter than its interaction length.
Formally, in thermal field theory cross sections get modified with respect to their leading-order in $g$
by effects suppressed by powers of $g/\pi$.
Thermal masses are a resummation of a class of such higher order corrections: those 
that become large at $E \circa{<} gT$.
Scatterings at higher order in $g$ can have extra initial state particles,
such as $V\chi\bar\chi \to B V$: this means that bound state formation is not blocked by thermal masses.
Technically, the same conclusion can be reached in the thermal formalism, by
computing the  formation rate of bound states $B$ rate as the imaginary part of their propagator $\Pi_{BB}$.
Cutted diagrams give an integral over thermal vectors: they have `poles' (that can get kinematically blocked) 
as well as `longitudinal'/`holes' and a `continuum' below the light cone,
which indeed corresponds to processes such as $V\chi\bar\chi \to B V$.

Formally, the cross section computed ignoring such `thermal mass' effects is correct at leading order in $g$.
In our cases of interest $g\sim g_3$ and $g \sim g_2$ are of order one, such that higher order effects cannot be neglected.
Given that a full thermal computation is difficult and does not seem to give
qualitatively new effects such as kinematical blocking of bound state formation,
{\em we compute the $\chi\bar\chi \to B V$ cross sections
at leading order in $g$ i.e.\ by ignoring the vector thermal mass $M_V$ in the kinematics}.
We take into account vector masses in the Yukawa potentials.
This approximation should be correct up to ${\cal O}(1)$ thermal corrections, as
confirmed by~\cite{LaineSommerfeld}, who finds that thermal corrections are small for $g=g_2$ and of order unity for $g=g_3$.

\section{Applications}\label{Applications}

We now apply our formalism to the computation of the thermal relic abundance of various models 
previously studied in the literature. We start with DM candidates with $\SU(2)$ quantum numbers, such us Minimal Dark Matter scenarios,
where the mass of mediators reduce the impact of bound state formation on the relic abundance. 
We then move to supersymmetric scenarios with co-annihilation of neutralinos with gluinos or squarks. 
We finally consider models  where a non-abelian gauge interaction dominates the annihilation cross-section.

\subsection{Minimal Dark Matter fermion triplet (wino)}\label{3plet}
The first explicit model that we consider is the Minimal DM fermionic triplet~\cite{Cirelli:2007xd}, which coincides
with a supersymmetric wino in the limit where all other sparticles are much heavier.
Once SU(2)$_L$ is broken, the conserved  quantum numbers are $L=0$, $S$ and $Q$.
The potential among the neutral states with spin $S=0$ is~\cite{HisanoCosmo,Cirelli:2007xd}
\beq \label{eq:S0triplet}
V_{Q=0}^{S=0}=\bordermatrix{&+&0\cr -&2\Delta M-A&-\sqrt{2}B \cr 0& -\sqrt{2}B&0 }.\eeq
where $A =   \alpha_{\rm em}/r + \alpha_2 c_{\rm W}^2 e^{-M_Zr}/r$, 
$B=\alpha_2e^{-M_Wr}/r$
and $\Delta M$ is the mass splitting produced by electroweak symmetry breaking,
equal to $\Delta M=165\MeV$ at $T=0$ (we use the two-loop result~\cite{0906.5207,1212.5989}).
The charged states with $S=0$ have~\cite{HisanoCosmo,Cirelli:2007xd}
\beq\label{eq:S+triplet}
V_{Q=1}^{S=0}=\Delta M+B ,\qquad
V_{Q=2}^{S=0}=2\Delta M+A .\eeq
Finally, for the states with $S=1$ one has
\beq\label{eq:S1triplet}
V_{Q=0}^{S=1} = 2\Delta M-A,\qquad
V_{Q=1}^{S=1}=\Delta M-B  .\eeq
where $V_{Q=1}^{S=1}$ differs by a sign from the earlier literature~\cite{HisanoCosmo,Cirelli:2007xd}.
These potentials allow to compute the Sommerfeld correction,
which affects the thermal relic abundance
because of the existence of a loosely bound state in the sector with $Q=S=0$ and $\ell=0$.
The cosmological DM abundance is reproduced for $\MDM\approx 2.7\TeV$,
such that the freeze-out temperature $\MDM/25$ is below the temperature at which $\SU(2)_L$ gets broken,
and the $\SU(2)_L$-invariant approximation is not accurate.

\begin{table}
\begin{center}
\begin{tabular}{c|c}
\rowcolor[HTML]{C0C0C0} 
& \\[-1.0ex]
\rowcolor[HTML]{C0C0C0} 
$I_{ J}\rightleftarrows I_{ J'}$ & $\sum\limits_{aMM'}|C_{\mathcal{J}}^{aMM'}+\gamma C_{\mathcal{T}}^{aMM'}|^2$   \\ [1.6ex] \toprule
$1\rightleftarrows 3$  & $2\left|1\mp\gamma \right|^2$                                    \\ \midrule
$3\rightleftarrows 5$  & $\frac{5}{2} \left|1\mp 2 \gamma\right|^2$                                \\ \midrule
\end{tabular}
\quad
\begin{tabular}{c|c}
\rowcolor[HTML]{C0C0C0} 
& \\[-1.0ex]
\rowcolor[HTML]{C0C0C0} 
$I_{J}\rightleftarrows I_{J'}$ & $\sum\limits_{aMM'}|C_{\mathcal{J}}^{aMM'}+\gamma C_{\mathcal{T}}^{aMM'}|^2$   \\ [1.6ex] \toprule
$1\rightleftarrows 3$  & $6\left|1\mp\gamma\right|^2$                                    \\ \midrule
$3\rightleftarrows 5$  & $\frac{21}{2} \left|1\mp2 \gamma \right|^2$                                \\ \midrule
$5\rightleftarrows 7$  & $12\left|1 \mp 3\gamma \right|^2$                      \\ \midrule
$7\rightleftarrows 9$  & $9\left|1\mp 4\gamma \right|^2$                      \\ \midrule
\end{tabular}
\end{center}
\caption{\em Group theory factors for formation of a bound state made of two $\SU(2)_L$ 3plets (left) or quintuplet (right)
with total isospin $I_{J'}$ from an initial state with total isospin $I_{J}$ and viceversa.
The upper sign refers to $I_J\to I_{J'}$, the lower sign to $I_{J'}\to I_J$.
\label{tab:C5}}
\end{table}

Nevertheless it is interesting to discuss the $\SU(2)_L$-invariant limit, which clarifies the controversial sign in eq.\eq{S1triplet}.
Ignoring $\SU(2)_L$ breaking, the DM-DM states formed by two triplets of $\SU(2)_L$ decompose in the following isospin channels
\begin{equation}
3\otimes 3 = 1_S \oplus 3_A\oplus 5_S\,,
\end{equation}
The two DM fermions can make a state with spin $S=0$ or 1.
The total wave function must be anti-symmetric under exchange of the two identical DM fermions:
taking into account the spin parity $(-1)^S$, the space parity $(-1)^\ell$ and the isospin parity
$(-1)^{\tilde I}$ where $I=2\tilde I+1$  is the dimension of the representation,
only states with $(-1)^{\ell+S+\tilde I} = 1$ are allowed.
Namely, the allowed states are
\be
\begin{tabular}{c|ccc|c}
$I$ & $V$ & \hspace{-3ex}\hbox{i.e.} \hspace{-3ex}& $\lambda$ & \hbox{allowed $\ell$}   \\  \hline
1 & $-2\alpha_2/r$ && $+2$ & \hbox{even if $S=0$, odd if $S=1$} \\ 
 3 & $-\alpha_2/r$ && $+1$ & \hbox{even if $S=1$, odd if $S=0$} \\ 
 5 & $+\alpha_2/r$ &&$-1$&  \hbox{even if $S=0$, odd if $S=1$}  \\ 
\end{tabular}
\label{3boundstates}
\ee
The charged components of the 5 two-body channel have potentials as in eq.\eq{S+triplet};
the neutral components of the $5$ mix with the 1 giving the matrix in eq.\eq{S0triplet}.
By computing its eigenvalues one finds that the correct $\SU(2)_L$-invariant limit is recovered for
$\Delta M=0$ and $A=B$.
The components of the $I=3$ triplet with $S=1$ have the potentials of eq.\eq{S1triplet},
with a correct $\SU(2)_L$-invariant limit:
notice that the $W$-mediated
$V_{Q=1}^{S=1}$ has opposite sign to $V_{Q=1}^{S=0}$, unlike what assumed in previous literature.
Anyhow, this channel is not attractive enough to form a bound state, so that the sign change has a minor impact,
as shown by comparing fig.~\ref{fig:F30}a with~\cite{HisanoCosmo,Cirelli:2007xd}. 
The figure also shows the DM abundance as obtained using the simple $\SU(2)_L$-invariant approximation, which
turns out not to be accurate.
In $\SU(2)_L$-invariant approximation the  Sommerfeld-corrected cross section~\cite{0806.1630} is obtained by decomposing  the total $s$-wave annihilation cross-section of eq.\eq{MDMtree}
into isospin channels:
\begin{equation}
\sigma_{\rm ann} v_{\rm rel} =\left[\frac {16}{111} S_2+ \frac {20}{111}  S_{-1} +\frac {75}{111}  S_1 \right]\times \frac {37}{12}\, \frac {\pi \alpha_2^2}{\MDM^2}.
\end{equation}
where $S$ is given by eq.\eq{Som} and the pedix  on $S$ indicates the value of $\lambda$.
We renormalise $\alpha_2$ at the RGE scale $M$, adopting the value from~\cite{1307.3536}.

\begin{figure}[t]
\begin{center}
\includegraphics[width=0.45\textwidth]{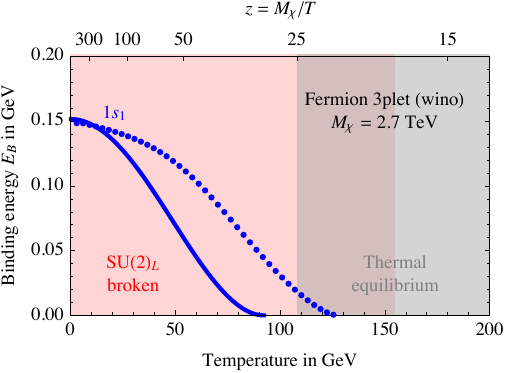}
\includegraphics[width=0.45\textwidth]{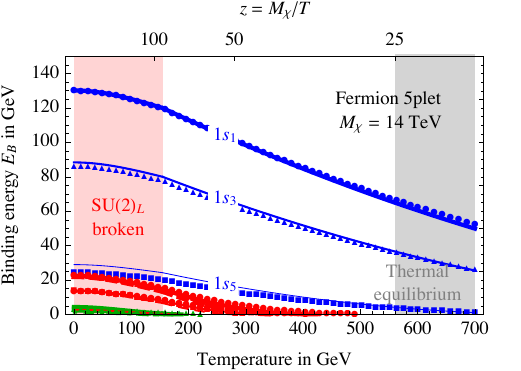}
\caption{\label{fig:EB5}\em  Energies of bound states at finite temperature.
made of two triplets  (left) or quintuplets (right).
Curves show results in $\SU(2)_L$-invariant approximation, dots show
numerical results in components.
In the left panel we consider DM as a $\SU(2)_L$ fermion triplet,
there is only one bound state and the $\SU(2)_L$-invariant approximation is not accurate.
In the right panel we consider DM as a $\SU(2)_L$ fermion quintuplet, 
the bound states are identified as follows:
$I=1$ (thick), $I=3$ (medium), $I=5$ (thin),
$n=1$ (blue), $n=2$ (red), $n=3$ (green), $\ell=0$ (continuous), $\ell=1$ (dashed), $\ell=2$ (dot-dashed).
}
\end{center}
\end{figure}

\begin{figure}[t]
\begin{center}
\includegraphics[width=.45\textwidth]{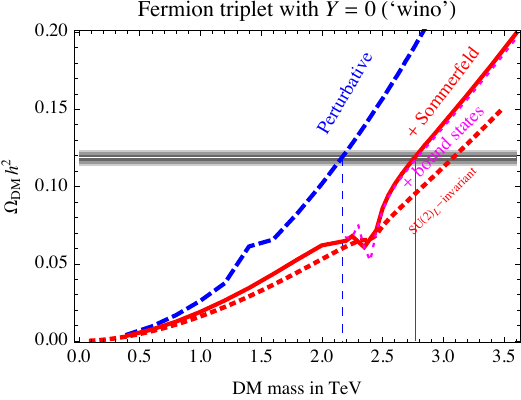}\qquad
\includegraphics[width=.45\textwidth]{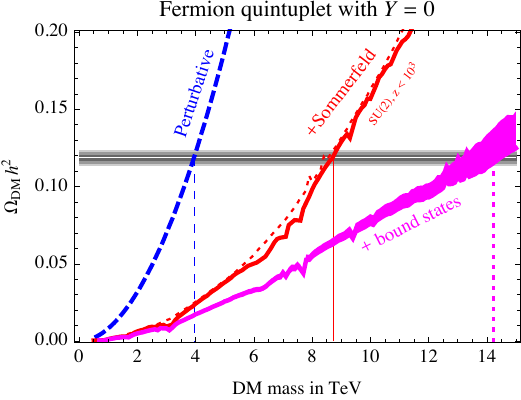}
\caption{\label{fig:quintuplet}\em Thermal relic DM abundance computed
taking into account tree-level scatterings (blue curve), adding Sommerfeld corrections (red curve),
and adding bound state formation (magenta).
We consider DM as a fermion $\SU(2)_L$ triplet (left panel)
and as a fermion quintuplet (right panel).
In the first case  the $\SU(2)_L$-invariant approximation is not good, but it's enough to show that
bound states have a negligible impact.
In the latter case the $\SU(2)_L$-invariant approximation is reasonably good, and adding
bound states has a sizeable effect.
\label{fig:F30}}
\end{center}
\end{figure}

\medskip

We next consider the contribution of bound states.
Eq.\eq{EboundYuk} tells that a bound state with given $n$ and $\alf=\lambda\alpha_2$ exists if
\beq \label{eq:BScond}\MDM \circa{>} 50 M_V \frac{n^2}{\lambda} \approx 4\TeV \frac{n^2}{\lambda}\eeq
where, in the last expression, we inserted the approximated vector mass  $M_W\approx M_Z$ at zero temperature.
This means that only the ground state $n=1$, $\ell=0$  of the $I=1$ configuration is present at $\MDM=2.7\TeV$,
in agreement with the component computation.
Thereby, we will consider only such $1s_1$ state (where the pedix denotes isospin).
Fig.\fig{EB5}a shows its binding energy as function of the temperature for $\MDM=2.7\TeV$.
The fact that the binding energy is small suggests that the Sommerfeld enhancement can be sizeable,
and that bound-state formation gives a small correction to the effective annihilation cross section.

The only existing bound state has $\ell=S=0$ and, in dipole approximation, can only be produced from an initial state with $\ell=1$
and $S=0$. No such state exists in the case of DM annihilations relevant for indirect DM detection,
where the initial state is  $\chi_0\chi_0$, that only exist with even $(-1)^{\ell+S}$ due to Pauli statistics~\cite{Slatyer}.
In the case of DM annihilations relevant for thermal freeze-out,
the bound state can be produced by $\chi_+\chi_-$ co-annihilations.
In the $\SU(2)_L$-invariant computation this difference arises because
we have isospin as an extra quantum number: the bound state with  $\ell=0$ and $I=1$
can be produced from an initial state with $\ell=1$, $I=3$. 
As discussed above, the $\SU(2)_L$-invariant approximation is not accurate; nevertheless it suffices to estimate
that the bound-state contribution is negligible.

Fig.~\ref{fig:EB5}a compares the approximated binding energy with the one
computed numerically from the full potential of eq.\eq{S0triplet}.
In $\SU(2)_L$-invariant approximation the annihilation width is $\Gamma_{\rm ann} = 8\alpha_2^5\MDM$, and the production cross section $\chi \chi \to B_{1s_1} \gamma$ is given by eq.\eq{bsf10} (with $C_{\cal J} = C_{\cal T} = \sqrt{2}$)
times  $\alpha_{\rm em}/3\alpha_2$ to take into account
that only the photon can be emitted (thermal masses do not kinematically block the process),
given that the non-thermal masses $M_{W,Z}$ are much bigger than the binding energy.
Even with this rough (over)estimate, bound-state formation affects the DM relic density by a negligible amount,
at the $\%$ level.
Its effect is not visible in fig.\fig{F30} where we show the DM thermal abundance as function of the DM mass.

\subsection{Minimal Dark Matter fermion quintuplet}\label{5plet}

\begin{figure}[t]
\begin{center}
\includegraphics[width=.85\textwidth]{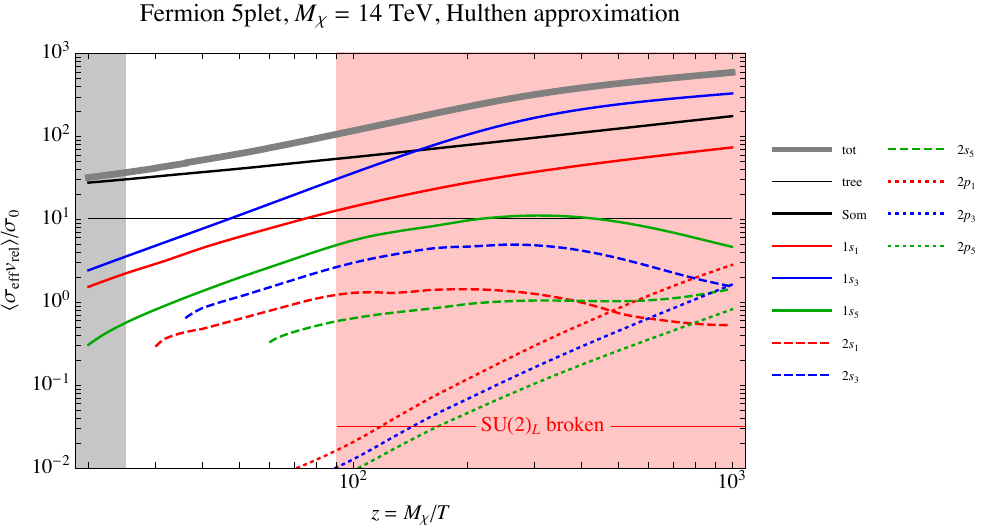}
\caption{\label{fig:quintuplet}\em  Assuming that DM is a fermionic $\SU(2)_L$ quintuplet,
we show its thermally-averaged
effective annihilation cross section at tree level in $s$-wave (horizontal line), 
adding Sommerfeld corrections (black curve), 
and the contributions from bound state formation for the bound states listed in eq.~(\ref{5boundstates}).}
\label{fig:quintuplet}
\end{center}
\end{figure}
We next consider the Minimal DM fermionic quintuplet~\cite{Cirelli:2007xd}. The DM-DM states formed by two quintuplets of $\SU(2)_L$ decompose into the following isospin channels
\begin{equation}
5 \otimes 5= 1_S \oplus 3_A \oplus 5_S \oplus 7_A \oplus 9_S.
\end{equation}
In the limit of unbroken $\SU(2)_L$ the $s$-wave annihilation cross-section reads~\cite{0706.4071}\footnote{Ref.~\cite{Cirelli:2007xd} 
performed a computation of Sommerfeld effects taking into account the breaking of $\SU(2)_L$.
In order to reproduce the correct $\SU(2)_L$-invariant limit, the non-abelian part of the potential in the sector with total electric charge
$Q=1$ and spin $S=1$ must be changed by a sign that makes it different from the sector with $Q=1$, $S=0$.
Eq.~(18) of~\cite{Cirelli:2007xd} must be changed into
\beq V_{Q=1}^{S=0} = 
 \bordermatrix{& ++ & +\cr
- & 5\Delta M  - 2 A & - \sqrt{6} B\cr
0 & -\sqrt{6} B&\Delta M + 3B},\qquad
V_{Q=1}^{S=1} = 
 \bordermatrix{& ++ & +\cr
- & 5\Delta M  - 2 A & - \sqrt{6} B\cr
0 & -\sqrt{6} B&\Delta M - 3B}.
\eeq
where
$A =   \alpha_{\rm em}/r + \alpha_2 c_{\rm W}^2 e^{-M_Zr}/r$ and 
$B=\alpha_2e^{-M_Wr}/r$
and $\Delta M$ is the mass splitting produced by electroweak symmetry breaking.
Namely, the sign of the non-abelian Coulomb potential depends on spin, unlike what assumed in earlier works.}
\begin{equation}
\sigma_{\rm ann} v_{\rm rel} = \frac{207}{20}
 \frac {\pi \alpha_2^2}{\MDM^2}\left[\frac {16}{69} S_6+ \frac {25}{69}  S_5 +\frac {28}{69}  S_3\right]\,
\end{equation}
where the tree-level cross section of eq.\eq{MDMtree}
has been decomposed into channels with $I=\{1,3,5\}$
(higher $I$ do not annihilate into SM particles), and the appropriate Sommerfeld
factor inserted for each channel.
The cosmological DM abundance is reproduced for $\MDM \approx 9.3\TeV$~\cite{0706.4071,Panci}.
Fig.~\ref{fig:F30}b  shows that the $\SU(2)_L$ invariant approximation can be reasonably good.
The approximation is exact at $T > T_{\rm cr}$, which includes the freeze-out temperature.
The approximations remains good below the critical temperature because electroweak vector masses are smaller than $\alf \MDM$,
and badly fails only at $T \circa{>} \Delta M$, when
the temperature gets smaller than the mass splittings $ \Delta M \sim \alpha_{\rm em} M_W$ between neutral and charged
DM components and co-annihilations become Bolztmann-suppressed.
In this temperature range $T \circa{>} M_W^2/M_\chi$, such that the Sommerfeld correction is well approximated
by its Coulombian limit.

\subsection*{Bound states}
In view of the selection rules discussed in the previous section, the allowed 
configurations are

\be
\begin{tabular}{c|ccc|c}
$I$ & $V$ & \hspace{-3ex}\hbox{i.e.} \hspace{-3ex}& $\lambda$ & \hbox{allowed $\ell$}   \\  \hline
1 & $-6\alpha_2/r$ && 6 & \hbox{even if $S=0$, odd if $S=1$} \\ 
 3 & $-5\alpha_2/r$ && 5 & \hbox{even if $S=1$, odd if $S=0$} \\ 
 5 & $-3\alpha_2/r$ &&3 &  \hbox{even if $S=0$, odd if $S=1$}  \\ 
 7 & 0 &&0 & \hbox{no bound state} \\ 
 9 & $4\alpha_2/r $&&$-4$ &  \hbox{no bound state} \\ 
\end{tabular}
\label{5boundstates}
\ee
where we have computed the non-abelian effective potential in each isospin channel.
Eq.\eq{BScond} 
shows that various bound states exist for $\MDM \sim 10\TeV$.
Taking thermal masses and the small dependence on $\ell$ into account, fig.\fig{EB5}b show the binding energies
as function of the temperature for $\MDM=14\TeV$.
We  consider formation of the $1s_I$, $2s_I$ and $2p_I$ `quintonium'
bound states in each isospin channel $I$:

\be
\begin{tabular}{ccccc|c|cc|c}
\hbox{Name}& $I$ & $S$ & $n$ & $\ell$ &$\lambda$  &$\Gamma_{\rm ann}/\MDM$  & $\Gamma_{\rm dec}/\MDM$ & \hbox{Produced from}\\  \toprule
$1s_1$ &$1$ & 0 & 1 & 0&6 &3240 $\alpha_2^5$ & 0 & $p_3$  \\ 
$1s_3$ & 3 & 1 & 1 & 0 &5 &  15625  $\alpha_2^5/48$ &0& $p_1$, $p_5$ \\ 
$1s_5$ & 5 & 0 & 1 & 0 &3 &  $567\alpha_2^5/4$ &0 & $p_3$, $p_7$  \\  \midrule
$2s_1$ &1 & 0 & 2 & 0&6 & 405  $\alpha_2^5$ & ${\cal O}(\alpha_2^4  \alpha_{\rm em}^2)$  &$ p_3$ \\
$2s_3$ & 3 & 1 & 2 & 0 &5 &  $15625  \alpha_2^5/384$ & ${\cal O}(\alpha_2^4  \alpha_{\rm em}^2)$ & $p_1,p_5$ \\
$2s_5$ & 5 & 0 & 2 & 0 &3 &  $567\alpha_2^5 /32$ & ${\cal O}(\alpha_2^4  \alpha_{\rm em}^2)$ & $p_3,p_7$ \\ \midrule
$2p_1$ & 1 & 1 & 2 & 1 &6 & ${\cal O}( \alpha_2^7)$ & $\approx 0.8\,\alpha_2^4 \alpha_{\rm em}$  & $s_3$ \\
$2p_3$ & 3 & 0 & 2 & 1 &5 & ${\cal O}( \alpha_2^7)$ & $\approx  0.5\, \alpha_2^4 \alpha_{\rm em}$ & $s_1,s_5$ \\
$2p_5$ & 5 & 1 & 2 & 1 &3 &   ${\cal O}( \alpha_2^7)$ &$\approx 0.2\,\alpha_2^4 \alpha_{\rm em}$ & $s_3,s_7$\\
\end{tabular}
\label{5boundstates}
\ee
The possibile initial states that can form each bound state are selected as follows.
In dipole approximation the value of the spin quantum number $S$ is conserved
and the angular momentum $\ell$ is changed by one unity.
Furthermore a vector boson is emitted, such that the initial isospin $I_{\rm in}$ must be $I \pm 2$.  
This leaves the possible initial states listed in the last column of the above table.

\smallskip

Each contribution to bound state formation is given by the generic formul\ae{} in section~\ref{bound} inserting the group theory factors
appropriate for the given $\SU(2)_L$ representations, as explicitly given in table~\ref{tab:C5}.
For example, let us consider the formation of the $1s_1$ bound state. 
The cross-section is given by eq.~(\ref{eq:bsf10}) and (\ref{eq:bsfC10}) with  
$
T_R=10$,
$d_R=5$,
$S=0$.
Once a bound state is formed, we need to determine its branching ratio into SM particles.
For $1s_I$ and $2s_I$ states, they are well approximated by eq.\eq{B1}.
For $2p_I$ states they are given by eq.\eq{2bs} and well approximated by
$\BR(2p_I \to 1s_I)\times \BR(1s_I\to \hbox{SM})$.

Fig.~\ref{fig:F30}b shows the DM cosmological abundance as function of its mass $\MDM$.
We summed the Sommerfeld-enhanced cross section (computed in $\SU(2)_L$ components)
with the bound-state  cross section computed in $\SU(2)_L$-invariant approximation.
As discussed above, the  $\SU(2)_L$ invariant approximation only holds at $T\circa{>}\Delta M$,
such that we switch-off  the bound-state contribution to the effective annihilation cross section
at $T < \MDM/10^3$ (upper border of the magenta band in fig.\fig{F30}b)
or at $T < \MDM/10^4$ (lower magenta band).
We adopted the couplings from~\cite{1307.3536} and normalized them at
$\MDM$ when computing annihilation rates, and at the inverse Bohr-radius 
$\alpha_2 \MDM$ when computing potentials.

We find that bound state formation increase by $\sim 40\%$ the effective annihilation cross section
defined in eq.\eq{YDMeff},  leading to a $\sim20\%$ increase in the value of $\MDM$ that reproduces
the cosmological DM abundance.
After including bound state formation, 
the cosmological DM abundance is reproduced for $\MDM \approx 14\TeV$.

\begin{figure}[t]
\begin{center}
\includegraphics[width=0.85\textwidth]{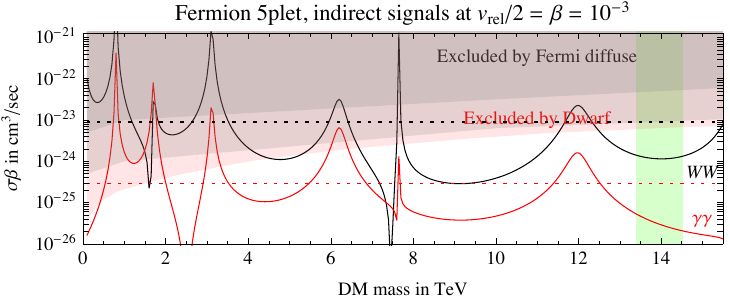}
\caption{\label{fig:quintupletCapture}\em 
Sommerfeld-enhanced cross section for indirect detection of a fermion 5-plet
at $\beta=10^{-3}$.
The grey areas are excluded by the Fermi diffuse bound (computed in a conservative way,
and in a more aggressive way); the red area is excluded by bound on dwarfs.
The dotted curves are the analytic $\SU(2)_L$-invariant approximation of eq.\eq{indSU2}.
}
\label{fig:quintupletind}
\end{center}
\end{figure}

\begin{figure}[t]
\begin{center}
\includegraphics[width=0.45\textwidth]{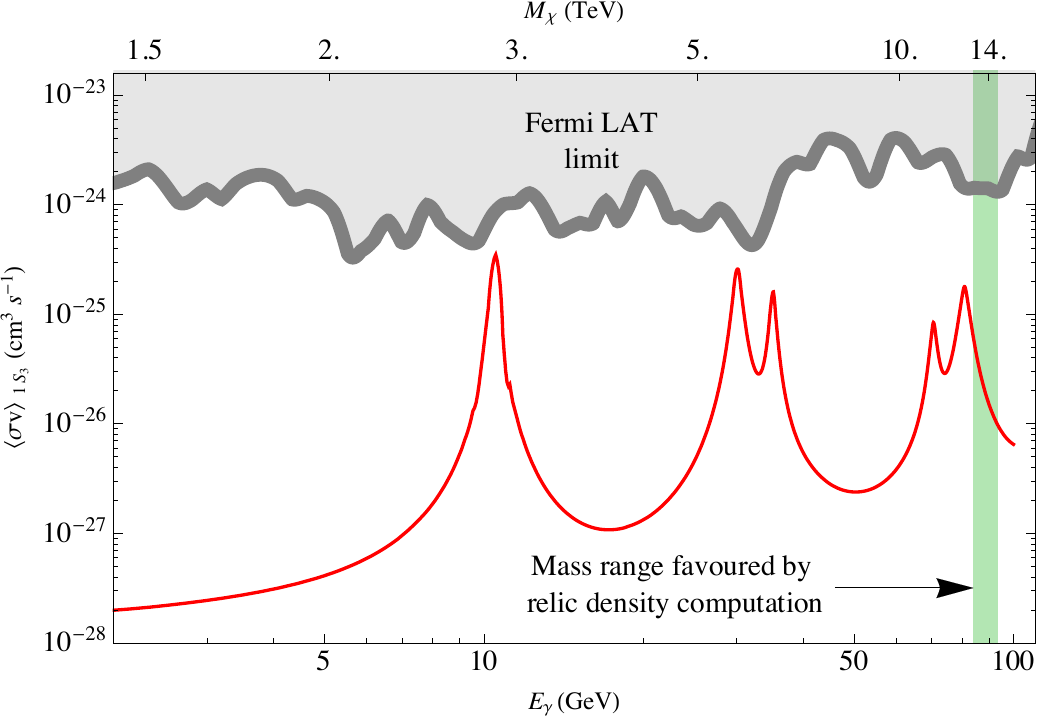}\qquad
\includegraphics[width=0.45\textwidth]{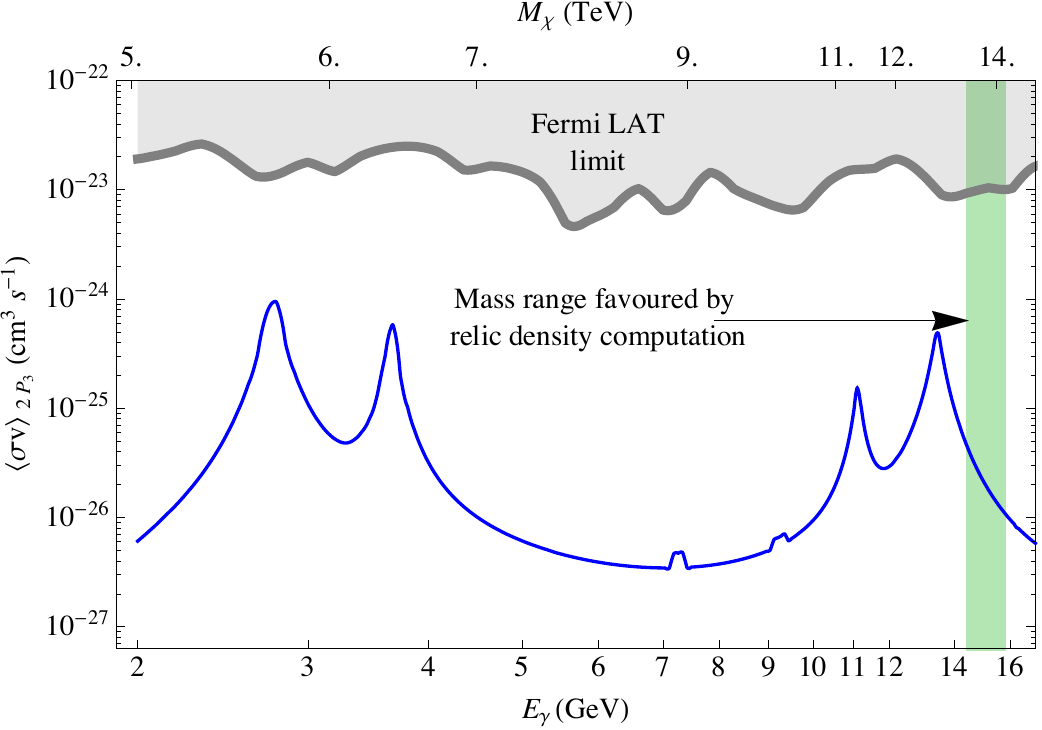}
\caption{\label{fig:quintupletCapture}\em 
Cross sections for producing a monochromatic photon after bound-state annihilation in the quintuplet model. 
We consider the contribution of the $1s_3$ (left) and $2p_3$ (right) bound state.
This signal cross section is compared with the bounds from  Fermi-LAT, assuming a contracted NFW DM density profile
and a  $3^\circ$ aperture around the galactic center (`R3' region)~\cite{Fermi2015}.  The Fermi-LAT limits on the $\gamma$-line cross sections have been appropriately
rescaled taking into account that one photon with energy smaller that the DM mass is emitted.}
\label{fig:quintuplet}
\end{center}
\end{figure}

\begin{figure}[t]
\begin{center}
\includegraphics[width=.45\textwidth]{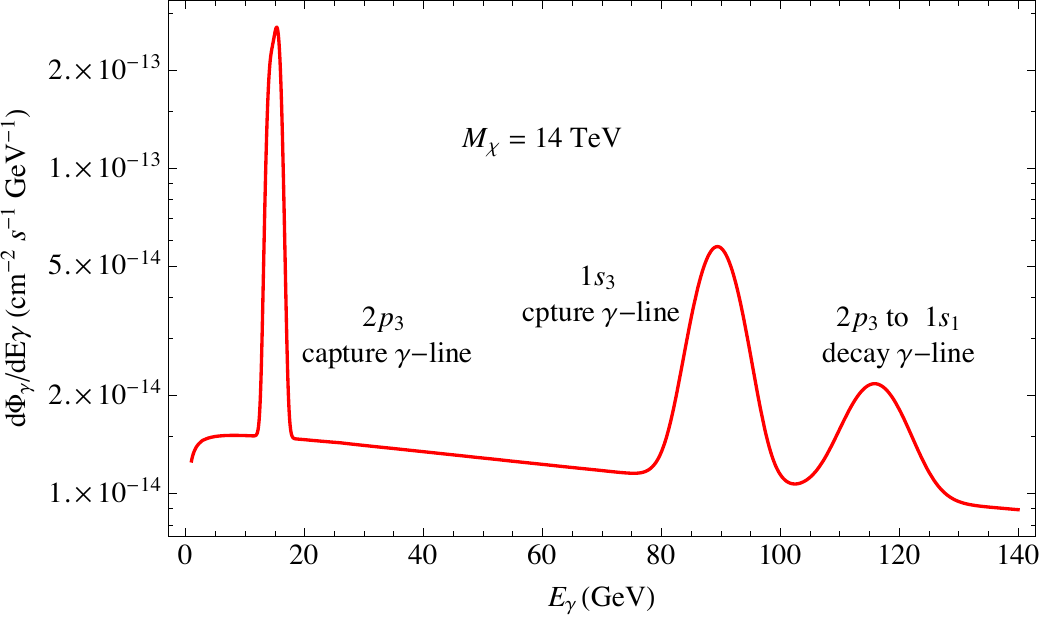}\qquad
\includegraphics[width=.42\textwidth]{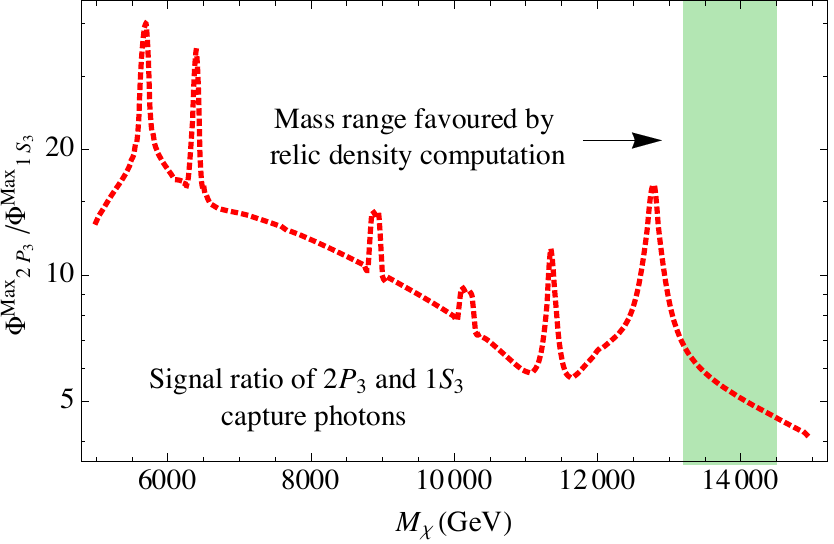}
\caption{\label{fig:quintupletRatio}\em In the left panel we show the $\gamma$-line spectrum predicted by the quintuplet model for the  $1s_3$ and $2p_3$ capture processes, computed in $\SU(2)_L$-invariant approximation.
A $10\%$ energy resolution of the  detector  is assumed. We choose a benchmark DM mass of  $\MDM=14\TeV$.  
In the right panel we show the ratio of the two  $\gamma$-line signal strength as a function of the DM mass. }
\label{fig:quintuplet}
\end{center}
\end{figure}

\subsection*{Indirect detection}
We now investigate the indirect detection prospects of the quintuplet dark matter model. A study of the direct annihilation of quintuplet dark matter leading to $W^+W^-$, $ZZ$, $\gamma\gamma$ has been performed in \cite{Cirelli:2007xd,Cirelli:2009uv,Panci,1507.05536,1608.00786}, finding
that the Sommerfeld enhancement plays a crucial role.
Photons resulting from $W,Z$ decays give a continuum photon spectrum,
which imply strong constraints if there is a large DM density around the Galactic Center,
see fig.~6 of~\cite{Cirelli:2007xd}.
However the DM density profile is unknown.
In fig.\fig{quintupletind} we compare the signal with the trustable
bound from the diffuse photon spectrum measured by {\sc Fermi}.
We show two bounds: a weaker safe bound obtained by demanding that the DM signal
(computed assuming a Burkert density profile)
never exceeds the measured spectrum, and a bound stronger by a factor $\approx 10$
obtained by subtracting the putative astrophysical background~\cite{Panci}.
The continuous curves is the prediction from a component computation~\cite{Cirelli:2007xd},
and the dotted curves are obtained from the $\SU(2)_L$-symmetric approximation
\beq\label{eq:indSU2}
(\sigma v_{\rm rel})_{WW} \approx \frac{8\pi \alpha_2^2}{\MDM^2} S_6 + \frac{2\pi\alpha_2^2}{\MDM^2} S_3,\qquad
(\sigma v_{\rm rel})_{\gamma\gamma} = \frac{4\pi \alpha_{\rm em}^2}{\MDM^2} S_6 + \frac{4\pi\alpha_{\rm em}^2}{\MDM^2} S_3
\eeq
as well as $\sigma_{ZZ}=\sigma_{\gamma\gamma}/\tan^4\theta_{\rm W}$, $\sigma_{\gamma Z} = 2\sigma_{\gamma\gamma}/\tan^2\theta_{\rm W}$.
Here $S_6$ ($S_3$) are the Sommerfeld factors for the $I=1$ and ($I=5$) channel:
around $M\approx 12\TeV$ they are enhanced by a zero-energy bound state with $n=4$ ($n=3$).
The formul\ae{}  above  correctly reproduce the peaks of the cross-section associated to zero energy bound states  while they 
miss the dips due to less important Ramsauer-Townsend effect, see \cite{1210.6104}.

Eq.\eq{indSU2} is obtained by writing the neutral component of the quintuplet as linear combination of
states with given total isospin,
\begin{equation}\label{eq:00}
|\chi_0 \chi_0\rangle = \frac 1 {\sqrt 5} |I=1,I_3=0\rangle - \sqrt{\frac 2 7}  |I=5,I_3=0\rangle +  \sqrt{\frac {18} {35}}  |I=9,I_3=0\rangle.
\end{equation}

\medskip

The continuum spectrum of photons resulting from $W,Z$ decays and fragmentations is not the most clean
experimental signal, given that astrophysics produce a largely unknown continuum background.
A monochromatic gamma line would give a clean signature, but a visible gamma line is not a generic feature of dark matter models~\cite{Duerr:2015vna,Duerr:2015aka}. 
We discuss here the possibility to search for quintuplet dark matter by looking for monochromatic 
photons emitted in the bound state formation processes $\chi_0\chi_0 \to B \gamma$.

The $\chi_0 \chi_0$ DM state of eq.\eq{00} can only exist with even ${\ell+S}$ due to Pauli statistics.
In the dipole approximation $\SU(2)_L$ conservation implies that only $I=3$ bound states can be formed
either from the $I=1$ or the $I=5$ component of $\chi_0\chi_0$:
the deepest such bound state is $1s_3$, with binding energy $E_B\approx 60\GeV (\MDM/10\TeV)$.
Therefore only the photon can be emitted in its formation, and consequently
only the neutral component of the bound state can be produced from $\chi_0\chi_0$.
The $M_{W,Z}$ masses cannot be neglected when computing the potentials.
Then, the cross section for bound state formation
is obtained by applying eq.s~(\ref{sys:sigmasgen})
to the desired single component, rather than summing over all possible components. The final result is
\beq
\sigma v_{\rm rel} (\chi_0\chi_0 \to B_{3n\ell} \gamma)=25 \frac{\alpha_{\rm em}}{\alpha_2}\bigg[
 \frac{1}{5} (\sigma v_\text{rel})^{n\ell}_\text{bsf} |_{\lambda_i=6,\lambda_f=5}^{C_{\cal J} = \sqrt{2},C_{\cal T} =-\sqrt{2}}
 +\frac{2}{7} (\sigma v_\text{rel})^{n\ell}_\text{bsf} 
 |_{\lambda_i=3,\lambda_f=5}^{C_{\cal J} = \sqrt{7/5},C_{\cal T} =+\sqrt{28/5}}
 \bigg]  \eeq
where the first (second) term corresponds to the $1\to 3$ ($5\to 3$) contribution, and the $I=3$ bound state $B$ is further identified by its
$n,\ell$ quantum numbers, and its spin is $S=1$ (0) for $\ell$ even (odd).
In this approximation we neglect the splitting between the various components of the multiplet. 
The above cross section is 2-3 orders of magnitude below $(\sigma v_{\rm rel})_{WW}$, with
a similar pattern of Sommerfeld enhancements: thereby the annihilation of the bound state into
$WW$ or similar states do nor produce relevant extra effects.
The interesting new feature is the monochromatic photon.

We average the cross section assuming that the DM velocity distribution in the galactic rest frame is a Maxwell-Boltzmann 
with root mean square velocity $220\,{\rm km/s}<v_0<270\,{\rm  km/s}$,
cut off by a finite escape velocity $450\,{\rm km/s}< v_{\rm esc}< 650\,{\rm  km/s}$:
\begin{equation}
\label{MB}
 f( v) = N\times  e^{-v^2/v_0^2} \, \theta(v_{\rm esc}-v) .
\end{equation}
The normalisation constant $N$ is fixed such that $\int d^3v~f(v) = 1$.
Furthermore we assume that all DM is made of 5plets.
We show the velocity-averaged photon capture cross sections in fig.~\ref{fig:quintupletCapture}.
The signal is below experimental bounds, and in some mass range it is close to the current sensitivity of the Fermi-LAT satellite.
Both lines from the $1s_3$ and the $2p_3$ capture processes appear to be in principle detectable in the future. 
Additionally, the $2p_3$ bound state decays into the $1s_1$ and $1s_5 $ states through emission of a photon, leading to extra gamma lines.  
In the Coulomb limit their energies are
\begin{equation}
E_{\gamma}=\frac {\alpha_2^2}{4} \left( \lambda_f^2- \frac{\lambda_i^2}4 \right) \MDM \,,
\end{equation}
where  $\lambda_i=5$ and $\lambda_f=(6,3)$ for $1s_1$ and $1s_5$ respectively and $\sfrac {\Gamma_{2p_3\to 1s_1+\gamma}}{\Gamma_{2p_3\to 1s_5+\gamma}}=0.38$.
This provides us with a  window of opportunity to obtain spectroscopic data about the dark matter in the universe and learn about its gauge interactions.  Fig.~\ref{fig:quintupletRatio}a shows that 
the extra peaks emerge over the continuum spectrum of photons from DM annihilations.
Fig.~\ref{fig:quintupletRatio}b shows  
 the ratio of the line signal intensities provides information about the dark matter mass. This information can then be confronted with searches for less specific emission of continuum photons at high energies stemming from direct dark matter annihilation.

\begin{figure}[t]
\begin{center}
\includegraphics[width=0.45\textwidth]{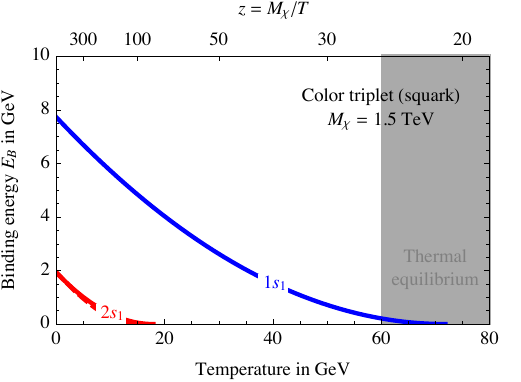}\qquad
\includegraphics[width=0.45\textwidth]{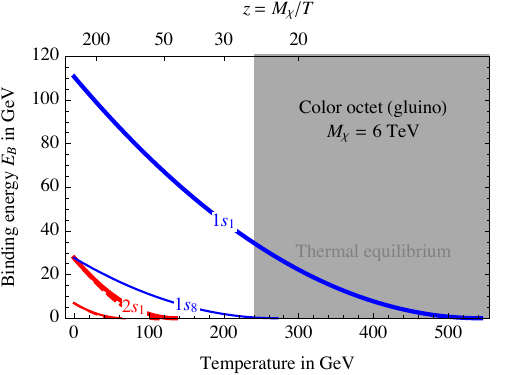}
\caption{\label{fig:EB3c}\em  Energies of bound states 
made of two squarks (left) or of two gluinos (right)
as color singlets (tick), color octets (thin),
$n=1$ (blue), $n=2$ (red), $\ell=0$ (continuous), $\ell=1$ (dashed).
}
\end{center}
\end{figure}

\subsection{Neutralino DM co-annihilating with a squark}\label{squark}
We next consider neutralino Dark Matter with mass close enough to a squark $\chi'=\tilde q$  such that co-annihilations determine
the relic abundance trough the effective cross section of eq.\eq{sigmav} as discussed in section~\ref{tree}.
The QCD process $\tilde q \tilde q^*\to gg$ dominates over
weak processes such as $\tilde q\tilde q \to qq$, that we neglect.
A squark $\tilde q$ is a scalar colour triplet, and a $\tilde{q} \tilde{q}^*$ state decomposes as $3\otimes \overline 3=1\oplus 8$.
The QCD potential $V = -\lambda_i\alpha_3/r$  is attractive with $\lambda_1=4/3$ in the singlet channel,
and repulsive with $\lambda_8=-1/6$ in the octet channel.
Squarks annihilate into gluons at tree-level in $s$-wave, and the 
cross section of eq.\eq{ann2gluons}
gets Sommerfeld-enhanced as~\cite{1402.6287}
\beq\sigma v_{\rm rel} = \frac{7}{27} \frac{\pi\alpha_3^2}{M_{\tilde q}^2}
\bigg[\frac{2}{7} S_{4/3} + \frac{5}{7} S_{-1/6}\bigg].\eeq
Bound states can exist in the color singlet channel with $S=0$ and any $\ell$, given that they are made of distinguishable scalars.
The lowest lying $\tilde q\tilde q^*$ bound states 
(we neglect $\tilde q \tilde q$ bound states, which can only annihilate trough weak processes) are
\be
\begin{tabular}{ccccc|c|cc}
\hbox{Name}& $R$  & $n$ & $\ell$ &$\lambda$  &$\Gamma_{\rm ann}/M_{\chi'}$  & $\Gamma_{\rm dec}/M_{\chi'}$ & \hbox{Produced from}\\  \toprule
$1s_1$ &$1$  & 1 & 0&4/3 &  $32\alpha_3^5/81$ & 0 & $p_8$  \\ 
$2s_1$ &$1$  & 2 & 0& 4/3 &  $4\alpha_3^5/81$ & $ {\cal O}(\alpha_3^6  )$  &$ p_{8}$ \\  
$2p_1$ & $1$  & 2 & 1 &4/3 & ${\cal O}( \alpha_3^7)$ & $O(\alpha_3^6) $  & $s_{8}$ \\
\end{tabular}
\label{squarkboundstates}
\ee
Taking into account the gluon thermal mass, and renormalizing the strong coupling at the inverse Bohr radius,
we find that the $1s_1$ bound state exists around the freeze-out temperature, see eq.\eq{lambdaBS}.
All other states only form at much lower temperatures,  as shown in fig.\fig{EB3c}a.
Even the binding energy of the $1s_1$ state gets significantly reduced by the gluon thermal mass,
indicating that the Coulomb approximation is not accurate.
We used the approximation described in section~\ref{sec:massive}.
The Clebsh-Gordon factors for bound-state formation are listed in table~\ref{tab:Ccolor}a.
Fig.\fig{coannBSF}a shows the
contribution of bound states to the total co-annihilation rate.
The contribution of the $1s_1$ state is 
accidentally suppressed 
because of a cancellation with the non-abelian contribution to gluon emission (last diagram in fig.\fig{bsf}),
The $2s_1$ state gives an order one reduction in the DM relic density, but only
at $T \sim \Lambda_{\rm QCD}$, when non-perturbative effects invalidate our computation.
We find that including bound states has a moderate impact on the DM relic density.

\medskip
\begin{table}[t]
\begin{center}
\begin{tabular}{c|c}
\rowcolor[HTML]{C0C0C0} 
& \\[-1.0ex]
\rowcolor[HTML]{C0C0C0} 
$R\rightleftarrows R'$ & $\sum\limits_{aMM'}|C_{\mathcal{J}}^{aMM'}+\gamma C_{\mathcal{T}}^{aMM'}|^2$   \\ [1.6ex] \toprule
$1\rightleftarrows 8$  &        $  \frac 4 3 |1\mp\frac 3 2\gamma |^2 $                        \\ \midrule
$3\rightleftarrows \overline{6}$  &          $  3|1\mp\gamma |^2 $           \\ \midrule
\end{tabular}
\quad
\begin{tabular}{c|c}
\rowcolor[HTML]{C0C0C0} 
& \\[-1.0ex]
\rowcolor[HTML]{C0C0C0} 
$R\rightleftarrows R'$ & $\sum\limits_{aMM'}|C_{\mathcal{J}}^{aMM'}+\gamma C_{\mathcal{T}}^{aMM'}|^2$   \\ [1.6ex] \toprule
$1_S\rightleftarrows  8_A$  & $3\left|1\mp \frac 3 2 \gamma \right|^2$                                    \\ \midrule
$8_A\rightleftarrows 8_S$  & 6                               \\ \midrule
$8_S\rightleftarrows 10_A\oplus \overline{10}_A$  & $3\left|2\mp3 \gamma\right|^2 $              \\ \midrule
$8_A\rightleftarrows 27_S$  & $9\left|1\mp\frac 5 2 \gamma\right|^2 $              \\ \midrule
\end{tabular}
\end{center}
\caption{\em Formation of a bound state made of two squarks (left) or two gluinos (right).
We show the group theory factors for formation of a bound state 
in the representation $R'$ made from an initial state in the representation $R$ and viceversa.
\label{tab:Ccolor}}
\end{table}

Furthermore, so far we have ignored the possibility that $\tilde q$ can decay, implicitly assuming that its life-time is long enough.
To conclude, we discuss what `long enough' means and whether this assumption is plausible.
A squark can decay into a neutralino DM and a quark, with rate
\begin{equation}\label{eq:Gammasquark}
\Gamma({\tilde q \to q  \chi}) \sim \frac {g^2} {8\pi}\frac {\sqrt{(M_{\tilde q}^2-\MDM^2)^2- 2 m_q^2(\MDM^2+M_{\tilde q}^2)+m_q^4}}{M_{\tilde q}}
\end{equation}
This new effect can be taken into account by the density-matrix formalism of eq.\eq{densitymatrix}, which
can be conveniently approximated by adding a stochastic term to the Schroedinger equation \eq{Schro},
represented by a non-unitary $\Gamma$ term in the Hamiltonian~\cite{quantumdiffusion}, such that
\begin{equation}
-\frac{\nabla^2\psi}{\MDM} +  V \psi = (E+ i\Gamma) \psi.
\end{equation}
As can be understood also from the uncertainty relation $\Delta E\, \Delta t >1$,
bound states only exist if the decay width $\Gamma$ is smaller than the binding energy $E_B \sim \alpha_3^2\MDMp$.
This is satisfied only if the squark decay width of eq.\eq{Gammasquark}
is strongly suppressed by the phase space.
Such kinematical suppression can reasonably happen if the squark is a stop $\tilde t$~\cite{0912.0526}, such that its tree-level 
decays into a top quark is kinematically blocked if 
$M_{\tilde t}- \MDM< M_t $,
allowing for a $\sim 5\%$ non-degeneration around $\MDM\sim 3\TeV$.
Furthermore,
at finite temperature this degeneracy gets broken by thermal corrections to the Higgs vev and to 
the squark mass $\Delta M_T \sim g^2_3 T^2/\MDM$, 
which effectively account for scatterings such as $g \tilde q\to \chi q$ that never get kinematically blocked,
giving rise to a thermal $\tilde q$ width $\Gamma \sim\alpha_3\alpha_1 T^3/\MDM^2$. 
Such effects can be neglected at the decoupling temperature $T_{\rm dec} \sim \MDM/25$.

\begin{figure}[t]
\begin{center}
\includegraphics[width=.48\textwidth]{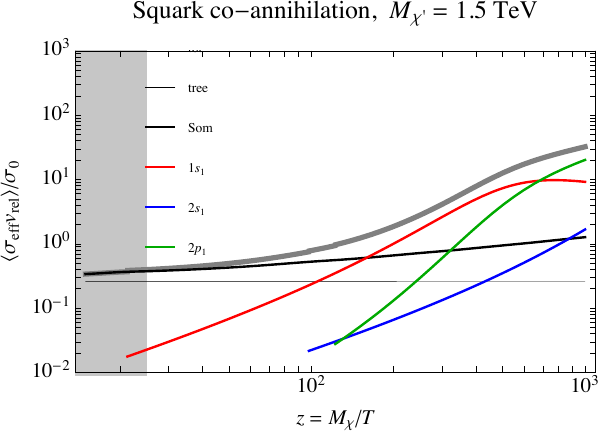}\quad
\includegraphics[width=.48\textwidth]{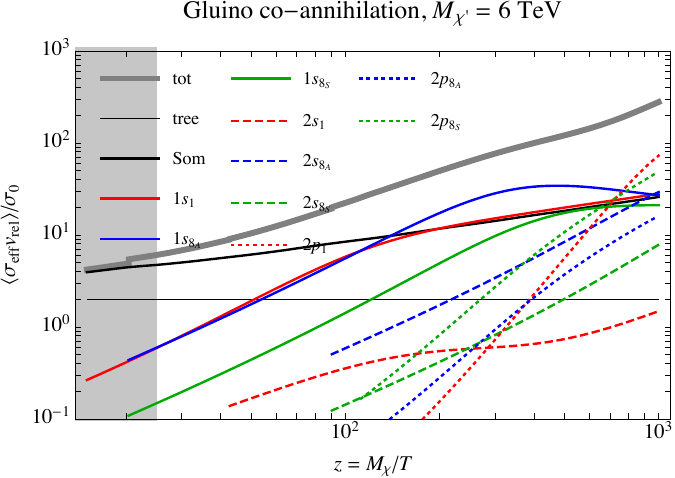}
\caption{\label{fig:coannBSF}\em  
Thermally-averaged
effective co-annihilation cross section at tree level in $s$-wave (horizontal line), 
adding Sommerfeld corrections (black curve), 
and the contributions from bound state formation for the bound states listed in eq.~(\ref{8boundstates}).}
\label{fig:gluinos}
\end{center}
\end{figure}

\subsection{Neutralino DM co-annihilating with a gluino}\label{gluino}
We next consider neutralino Dark Matter with mass close enough to a gluino $\tilde g$  such that co-annihilations determine
the relic abundance  through the effective cross section of eq.\eq{sigmav}.
The product of two color octets decomposes as
\begin{equation}
8 \otimes 8= 1_S \oplus 8_A \oplus 8_S \oplus 10_A \oplus \overline{10}_A\oplus 27_S .
\end{equation}
Each channel experiences the following potentials
\be
\begin{tabular}{c|ccc|c}
\hbox{Color} & $V$ & \hspace{-3ex}\hbox{i.e.} \hspace{-3ex}& $\lambda$ & \hbox{allowed $\ell$}   \\  \hline
$1_S$ & $-3\alpha_3/r$ && 3 & \hbox{even if $S=0$, odd if $S=1$} \\ 
 $8_A$ & $-\frac 3 2 \alpha_3/r$ && $\sfrac 3 2$& \hbox{even if $S=1$, odd if $S=0$} \\ 
 $8_S$ & $-\frac 3 2 \alpha_3/r$ && $\sfrac 3 2$ &  \hbox{even if $S=0$, odd if $S=1$}  \\ 
 $10_A\oplus \overline{10}_A$ & 0 &&0 & \hbox{no bound state} \\ 
 $27_S$ & $\alpha_3/r $&&$-1$ &  \hbox{no bound state} \\ 
\end{tabular}
\label{5boundstates}
\ee
where on the last column we listed the bound states supported in the attractive channels.
The symmetric channels can annihilate into two gluons at tree level, and the $8_A$ channel can annihilate into quarks:
the Sommerfeld-corrected $s$-wave annihilation cross-section is~\cite{1402.6287}
\begin{equation}
\sigma v_{\rm rel} =\frac {27} {32}\sigma_0 \left[\frac 1 6 S_{3}+\frac 1 3 S_{ 3 /2}+ \frac 1 2  S_{-1}\right]+ 
 \frac {9} {8} \sigma_0  S_{3/2},\qquad \sigma_0=\frac{\pi \alpha_3^2}{M_{\tilde g}^2}
\end{equation}
where the first (second) term comes from annihilations into gluons (quarks).


\begin{figure}[t]
\begin{center}
\includegraphics[width=0.46\textwidth]{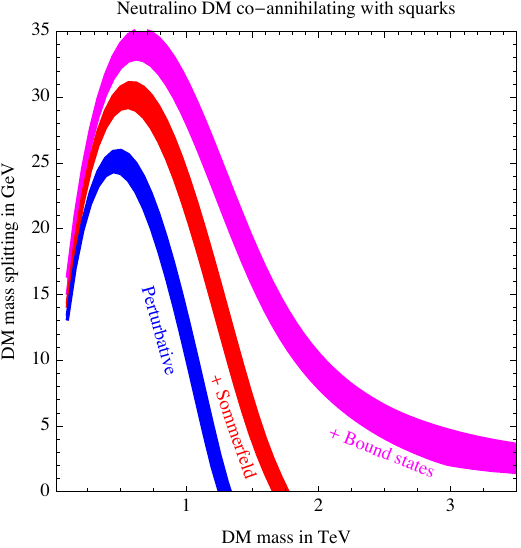}\qquad
\includegraphics[width=0.46\textwidth]{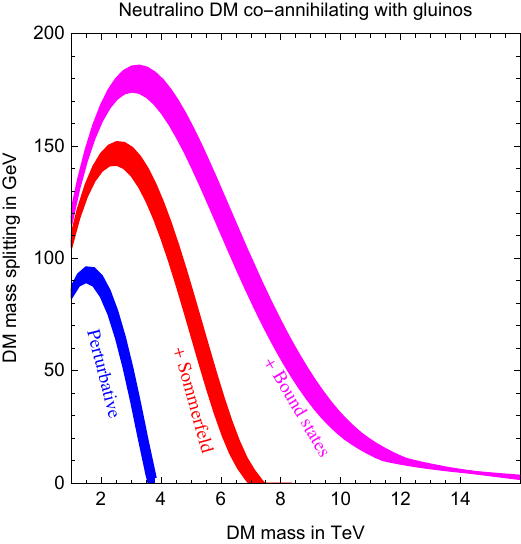}\caption{\label{fig:gluinoCAN}\em The colored bands represent the regions in the plane of mass splitting between the colored partnter (gluino/squark)  and the dark matter (neutralino) in which the correct relic abundance is reproduced within three standard deviations. The computation has been performed at tree-level (blue), taking into account Sommerfeld enhancement (red) and bound state formation (magenta).
In the latter case, the tail at large $M$ and small $\Delta M \sim \Lambda_{\rm QCD}$ is due to QCD effects enhanced by confinement, 
and more in general by the large value of the strong coupling, following~\cite{1801.01135}.
}
\label{fig:gluinoCAN}
\end{center}
\end{figure}

Furthermore,  around the freeze-out temperature
two (one) bound state in the singlet (octet) channel exist, 
as illustrated in fig.\fig{EB3c}b, which takes the gluon thermal mass into account.
We assume that gluino decay is slow enough, $\Gamma_{\tilde{g}}\ll E_B$, that  gluino bound states can form.
Furthermore, we assume that the gluino and DM are kept in relative equilibrium.
If DM is a bino, these assumptions are satisfied by having relatively heavy squarks.
Gluino bound states have been considered in~\cite{Ellis:2015vaa,1611.08133} where the gluon thermal mass
was neglected and only the singlet bound states with $n=1$ was included. 
Furthermore, we include the non-abelian contribution to bound-state formation (latter diagram in fig.\fig{bsf}), whose effect is described by the $C_{\cal T}$ contribution in table~\ref{tab:Ccolor}b,
and our  $C_{\mathcal{J}}$ differs by a factor $1/\sqrt{2}$.

At zero temperature the lowest lying bound states are:
\be
\begin{tabular}{ccccc|c|cc|c}
\hbox{Name}& $R$ & $S$ & $n$ & $\ell$ &$\lambda$  &$\Gamma_{\rm ann}/\MDM$  & $\Gamma_{\rm dec}/\MDM$ & \hbox{Produced from}\\  \toprule
$1s_1$ &$1_S$ & 0 & 1 & 0&3 &  $243\alpha_3^5/4$ & 0 & $p_{8_A}$  \\ 
$1s_{8_A}$ & $8_A$ & 1 & 1 & 0 &3/2 &   $81\alpha_3^5/32 $ &0& $p_1$, $p_{8_S}$, $p_{27_S}$ \\ 
$1s_{8_S}$ & $8_S$ & 0 & 1 & 0 &3/2 &  $243\alpha_3^5/128$ &0 & $p_{8_A}$, $p_{10_A}$  \\  \midrule
$2s_1$ &$1_S$ & 0 & 2 & 0&3 &  $243\alpha_3^5/32$ & $ {\cal O}(\alpha_3^6  )$  &$ p_{8_A}$ \\
$2s_{8_A}$ & $8_A$ & 1 & 2 & 0 &3/2 &  $81 \alpha_3^5/256 $ & ${\cal O}(\alpha_3^6 )$ & $p_1,p_{8_S}$, $p_{27_S}$ \\
$2s_{8_S}$ & $8_S$ & 0 & 2 & 0 &3/2 &  $243\alpha_3^5/1024$ & ${\cal O}( \alpha_3^6 )$ & $p_{8_A},p_{10_A}$ \\ \midrule
$2p_1$ & $1_S$ & 1 & 2 & 1 &3 & ${\cal O}( \alpha_3^7)$ & $\approx\,\alpha_3^6 $  & $s_{8_A}$ \\
$2p_{8_A}$ & $8_A$ & 0 & 2 & 1 &3/2 & ${\cal O}( \alpha_3^7)$ & $\approx   0.1 \alpha_3^5 $ & $s_1,s_{8_S}$, $s_{27_S}$ \\
$2p_{8_S}$ & $8_S$ & 1 & 2 & 1 &3/2 &   ${\cal O}( \alpha_3^7)$ &$\approx \,0.1 \alpha_3^5$ & $s_{8_A},s_{10_A}$\\
\end{tabular}
\label{8boundstates}
\ee
Fig.\fig{gluinos}b shows how each bound state contributes to the effective annihilation cross section, and
fig.\fig{gluinoCAN}b shows how the resulting DM abundance gets affected.
We find a moderate shift of the regions where the thermal abundance reproduces the cosmological DM abundance.
The largest effect arises when  $ M_{\tilde g} - \MDM$ is small, such that 
formation of $2p$ bound states from $s$-wave free states become sizeable at low temperatures.

\section{Conclusions}\label{concl}
In the first part of the paper we presented generic expressions and tools for computing non-abelian bound state formation.
We specialised these formul\ae{} to an unbroken gauge group, such that a significant
simplification over a component computation is obtained making use of group algebra.
We applied these results to study how formation of bound states of two Dark Matter particle decrease their thermal abundance,
in various concrete DM models.
\begin{enumerate}
\item In section~\ref{3plet} we assumed that Dark Matter is a fermionic 3plet of $\SU(2)_L$ with zero hypercharge,
for example a supersymmetric wino. 
We find that the $\SU(2)_L$-invariant approximation is only qualitatively accurate.
Nevertheless it is enough to establish that bound states have a negligible impact, at the $\%$ level, on the thermal relic DM abundance.
Furthermore, it shows that the non-abelian Coulomb energy depends on total spin, 
unlike what assumed in previous computations: we thereby repeat a component computation with the correct signs,
and including thermal corrections to the weak mass splitting between charged and neutral components of the DM multiplet.

\item In section~\ref{5plet} we assumed that Dark Matter is an automatically stable 
fermionic 5plet of $\SU(2)_L$ with zero hypercharge.
We found that `quintonium' bound states reduce the DM thermal abundance by about $30\%$,
increasing the DM mass that reproduces the cosmological abundance to about $14\TeV$. 
We also considered bound-state corrections to DM indirect detection, finding that
the 5-plet predicts a characteristic spectrum of mono-chromatic $\gamma$ lines around $E_\gamma \sim (10-80)\GeV$,
with rates of experimental interest.

\item 
In section~\ref{squark} we have considered Dark Matter co-annihilating with a scalar color triplet, a squark in supersymmetric models,
finding that bound state give  a mild shift in the thermal relic density.

\item 
In section~\ref{gluino} we have considered Dark Matter co-annihilating with a fermionic color octet, a gluino in supersymmetric models,
improving the results of~\cite{Ellis:2015vaa} by taking into account thermal masses and bound-state formation
with gluon emission form gluons, as depicted in the last diagram of fig.\fig{bsf}.
Bound state formation gives a significant correction to the thermal relic DM density.

\end{enumerate}
We think that our results should be improved along two lines.
First, concerning the weak 5plet,
a computation in components will be needed for a precision computation that takes into account that  $\SU(2)_L$ is broken.
Second, we have taken into account thermal masses, and assumed that they do not kinematically block bound-state formation
for the reasons discussed in section~\ref{finiteTbsf}.  While we expect this to be a reasonable approximation,
a  careful study of thermal effects, possibly along the lines of~\cite{LaineSommerfeld},
will be needed to achieve a more precise result.

 \subsubsection*{Acknowledgments}
We thank Francesco Becattini, Martin Beneke, 
Camilo Alfredo Garcia Cely, Marco Cirelli, Paolo Panci, Filippo Sala, Tracy Slatyer for discussions.
This work was supported by the ERC grant NEO-NAT and  by the  MIUR-FIRB grant RBFR12H1MW.

\appendix

\small

\section{Wave functions in a potential mediated by a vector}\label{app:crosssect}

In this appendix we collect the relevant formulas used throughout the paper.

If the vector is massless, the radial wave functions of a bound state in the Coulomb potential are
\beq \label{eq:psiCoulomb}
R_{n\ell}(r) =   \bigg(\frac{2}{na_0}\bigg)^{3/2}
\sqrt{\frac{(n-\ell-1)!}{2n(n+\ell)!}} e^{-r/na_0} \bigg(\frac{2r}{na_0}\bigg)^\ell
L_{n-\ell-1}^{2\ell+1}(\frac{2r}{na_0})
\eeq
where $a_0= 2/\alf \MDM$ is the Bohr radius and $L$ are Laguerre polynomials.
If the vector has mass $M_V$, an analytic solution is obtained
approximating the Yukawa potential with a  Hulthen potential
\begin{equation}
V_{\rm Hulthen}= \frac{  \kappa\, M_V e^{- \kappa M_V r }}{1- e^{-\kappa\, M_V r}}.
\end{equation}
The radial wave functions of bound states in the Hulthen potential are
\beq R_{n\ell}(r) = N_{n\ell}   e^{-r \kappa M_V q_{n\ell} }    \frac{(1-e^{-r\kappa  M_V})^{\ell+1}}{r}
 P_{n-\ell-1}^{ 2q_{n\ell}, 1+2\ell }
(1-2 e^{-r\kappa  M_V}) 
   \eeq
where 
$ q_{n\ell} = \sfrac{\sqrt{\MDM E_{n\ell}}}{\kappa M_V} $,  
$N_{n\ell}$ is the normalization factor such that $\int dr\,r^2 R_{n\ell}R_{n'\ell}=\delta_{nn'}$,
and  $P$ are the Jacobi polynomials\footnote{Implemented in {\tt Mathematica} as 
 $P^{b,c}_a(x)={\tt JacobiP[a,b,c,x]}$.
 The value of $c$ differs from~\cite{Slatyer}. }
which equal unity for $\ell = n-1$.
For $\ell=0$ one has
 \beq q_{n0} = \frac{1-n^2  y}{2n y},\qquad
 N_{n0} =\sqrt{\kappa M_V \frac{1-n^4 y^2}{2y^3 n^5}}.\eeq
In particular, the ground-state wave function is
 \beq R_{10}(r) = \sqrt{\kappa M_V \frac{1-y^2}{2y^3}}
 e^{-r \kappa M_V q_{10}}  \frac{1-e^{-r\kappa  M_V}}{r}.
\eeq
The normalisation factor for $\ell=1$ is
$
N_{n1} = N_{n0}\sqrt{\sfrac{(1 - n^2y^2) }{(n^2-1)n^2y^2}}
$.

\bigskip

The normalized radial wave function of a free state in the Hulthen potantial is~\cite{Cassel:2009wt,Slatyer}
\beq \begin{array}{rcl}R_\ell(r) &=& \displaystyle
\sqrt{\frac{4\pi}{2\ell+1}}  \frac{(1-e^{-\kappa M_V r})^{\ell+1}}{\MDM r} 
e^{-i \MDM v_{\rm rel} r/2} \,
{}_2F_1(a^-,a^+,2(\ell+1),1-e^{-\kappa M_V r})\times\\
&&\times\displaystyle{\frac{1}{(2\ell)!}} 
\frac{\sqrt{S}}{\alf}\prod_{k=0}^\ell \sqrt{(\frac{\alf\MDM}{\kappa M_V})^2+k^2(\frac{v_{\rm rel}\MDM}{\kappa M_V})^2(1-\frac{2\kappa M_V\alf}{\MDM v_{\rm rel}^2})+k^4}
\end{array}
 \eeq
where $S$ is the Sommerfeld factor for $\ell=0$ given in eq.\eq{Som},
$F$ is the hypergeometric function\footnote{Implemented in {\tt Mathematica} as 
 $_2 F_1(a,b;c;x)={\tt Hypergeometric2F1[a,b,c,x]}$.} 
and its arguments are
\beq a^\pm = 1+\ell+i \MDM\frac{v_{\rm rel}}{2\kappa M_V} \bigg(1 \pm \sqrt{1-\frac{4\kappa M_V\alf}{\MDM v_{\rm rel}^2}}\bigg).\eeq
The function $R_\ell(r)$ is real, and
in the limit $\alf=0$ it reproduces the free partial wave expansion
$ e^{i \vec r \cdot \vec p} = \sum_\ell i^\ell R_\ell(r)Y_{\ell 0}(\theta)$ where 
$R_\ell(r)= \sqrt{4\pi(2\ell+1)} i^\ell j_\ell(pr)$.
Here $\theta$ is the angle between $\vec r$ and $\vec p$,
$p = \MDM v_{\rm rel}/2$ and 
$j_\ell$ are spherical Bessel functions
$j_\ell(z) = \sqrt{\pi/2z} J_{\ell+1/2}(z)$
(equal to  $j_0(z) \simeq \sin(z)/z$ for large $z$).
Furthermore, in the massless limit $M_V=0$, $R_\ell$ reproduces the Coulomb partial wave expansion
\beq \label{eq:FreeCoulomb}
e ^{\pi  \alf/2v_{\rm rel}} \Gamma( 1-i \alf/v_{\rm rel})  \,
{}_{1}F_1 \left[ i \alf/v_{\rm rel} , 1, i (p r- \vec p\cdot \vec r) \right] e^{i \vec r \cdot \vec p} 
 = \sum_\ell i^\ell  R_\ell(r)  Y_{\ell 0}(\theta) 
 \eeq
where 
\beq
R_\ell (r)=  
\frac { \sqrt{4\pi(2\ell+1) \,S} }{\Gamma(2\ell+2)}e^{-i p r} (2 p r)^\ell ~_1F_1[\ell+1+i \alf/v_{\rm rel},2\ell+2,2 i p r]   \prod_{k=1}^\ell \left|\ell -k + 1 - i \alf/v_{\rm rel} \right|  .\eeq
 Such analytic solution of the wave function for the Hulthen potential is only exact if $\ell =0$. 
 If $\ell$ is not zero, an extra approximation for the centrifugal term is needed, such that the behaviour at large $r$ becomes only approximate. 
 This is not a problem  for the bound state wave function, as it is exponentially suppressed at large $r$. 
 However, for the free state this approximation leads to unphysical results that become relevant in the case of bound state production from a $p$-wave and $d$-wave partial waves. In order to correct for this inaccuracy we multiply the resulting cross sections, as was suggested in~\cite{Cassel:2009wt} by
\beq
 L_\ell =  \frac{w^{2 \ell}}{\prod_{k=0}^{k=\ell-1}\left((\ell -k)^2 +w^2 \right) } 
 \qquad\text{  with  } \qquad
 w =\frac{ \MDM \,v_\text{rel}}{ \kappa M_V} \approx \frac{p}{M_V}\,. 
\eeq
This function is controlled by the critical momentum $M_V$ so that, once the momentum of the dark matter particles drops below the mediator mass,  the production cross sections from higher $\ell$ states are suppressed.

\end{document}